\documentclass[a4paper,structabstract]{aa}
\pdfoutput=1
 \usepackage{mathtools} 
\usepackage{bbding}
\usepackage{mathenv}
\usepackage{aas_macros}
\usepackage{natbib}
\usepackage{txfonts}
\usepackage{array}
\usepackage{wasysym} 
 \usepackage[squaren,grey]{SIunits}
 \usepackage{enumerate}

\usepackage{graphicx}

 \usepackage{booktabs} \newcommand{\ra}[1]{\renewcommand{\arraystretch}{#1}}

\def\wig#1{\mathrel{\hbox{\hbox to 0pt{%
          \lower.6ex\hbox{$\sim$}\hss}\raise.4ex\hbox{$#1$}}}}

\def\ie{{\it i.e.}}
\def\eg{{\it e.g.}}

\def\teff{T_{\rm eff}}

\def\tint{T_{\rm int}}
\def\tirr{T_{\rm irr}}
\def\teffmu{T_{\rm eff,\,\mu_{*}}}

\def\tskin{T_{\rm skin}}

\def\fh{f_{\rm H}}
\def\tdeep{T_{\rm deep}}

\def\Jv{J_{\rm v}}
\def\Hv{H_{\rm v}}

\def\Ka{\kappa_{\rm 1}}
\def\Kb{\kappa_{\rm 2}}
\def\Kp{\kappa_{\rm P}}
\def\Kr{\kappa_{\rm R}}

\def\Knu{\kappa_{\nu}}

\def\Rt{R}

\def\taulim{\tau_{\rm lim}}
\def\Ja{J_{\rm 1}}
\def\Jb{J_{\rm 2}}

\def\Ga{\gamma_{\rm 1}}
\def\Gb{\gamma_{\rm 2}}
\def\Gp{\gamma_{\rm P}}
\def\Gv{\gamma_{\rm v}}
\def\Gab{\gamma_{(\rm 1, \rm 2)}}

\def\Ca{C_{1}}
\def\Cb{C_{2}}
\def\Cc{C_{3}}
\def\Cd{C_{4}}

\def\D{\mathrm{d}}

\def\JG{J_{\rm \gamma}}
\def\JGa{J_{\rm \gamma}}
\def\JGb{J_{\rm \gamma^{3}}}
\def\Kv{\kappa_{\rm v}}
\def\Ha{H_{\rm 1}}
\def\Hb{H_{\rm 2}}

\def\Knu{\kappa_{\nu}}

\usepackage{hyperref}
\hypersetup{
colorlinks=true,
citecolor=blue,
linkcolor=blue,
urlcolor=black
}

\begin{document}

\title{A non-grey analytical model for irradiated atmospheres.}
\subtitle{I: Derivation}
\titlerunning{A non-grey analytical model for irradiated atmospheres}
\authorrunning{V. Parmentier \& T. Guillot}

\author{Vivien Parmentier\inst{1,2}
\and
Tristan Guillot\inst{1,2}}

\offprints{V.Parmentier}

\institute{Laboratoire Lagrange, UMR7293, Universit\'e de Nice Sophia-Antipolis, CNRS, Observatoire de la C\^ote d'Azur, 06300 Nice, France \email{vivien.parmentier@oca.eu}\\
\and 
Department of Astronomy and Astrophysics, University of California, Santa Cruz, CA 95064, USA
}

\date{Re-submitted to A\&A \today}

\abstract
{Semi-grey atmospheric models (with one opacity for the visible and one opacity for the infrared) are useful to understand the global structure of irradiated atmospheres, their dynamics and the interior structure and evolution of planets, brown dwarfs and stars. But when compared to direct numerical radiative transfer calculations for irradiated exoplanets, these models systematically overestimate the temperatures at low optical depth, independently of the opacity parameters.}
{We wish to understand why semi-grey models fail at low optical depths, and provide a more accurate approximation to the atmospheric structure by accounting for the variable opacity in the infrared.}
{Using the Eddington approximation to link the energy of the beam to its radiation pressure, we derive an analytical model to account for lines and/or bands in the infrared. Four parameters (instead of two for the semi-grey models) are used: a visible opacity $\kappa_{\rm v}$, two infrared opacities, $\kappa_{\rm 1}$ and $\kappa_{\rm 2}$, and $\beta$ the fraction of the energy in the beam with opacities $\kappa_{\rm 1}$. We consider that the atmosphere receives an incident irradiation in the visible with an effective temperature $T_{\rm irr}$ and at an angle $\mu_{*}$, and that it is heated from below with an effective temperature $T_{\rm int}$.  }
{Our irradiated non-grey model is found to provide a range of temperatures that is consistent with that obtained by numerical calculations. We find that because of the variable opacities in the infrared, a smaller fraction of the source function is contributing to the heating of the upper layers leading to smaller temperatures than in the semi-grey models. For small values of $\beta$ (expected when lines are dominant), we find that the non-grey effects are confined to low-optical depths. However, for $\beta\wig{>} 1/2$ (appropriate in the presence of bands with a wavelength-dependence smaller or comparable with the width of the Planck function), we find that the temperature structure is affected even down to infrared optical depths unity and deeper as a result of the so-called blanketing effect. }
{The expressions that we derive may be used to provide a proper functional form for algorithms that invert the atmospheric properties from spectral information. Because a full atmospheric structure can be calculated directly, these expressions should be useful for simulations of the dynamics of these atmospheres and of the thermal evolution of the planets. Finally, they should be used to test full radiative transfer models and improve their convergence. }


\maketitle
%

\section{Introduction}

The discovery of numerous star-planet systems and the possibility to characterize the planets' atmospheric properties has led to a great many publications using radiative transfer calculations, often ``off-the-shelf'' from numerical models. Given the a priori infinite amount of possible compositions for mostly unknown exoplanetary atmospheres, it is highly valuable to have the possibility to perform very fast calculations and also understand what determines the thermal structure of an irradiated atmosphere.

Analytical radiative transfer solutions for atmospheres have been calculated with a variety of assumptions and different contexts \citep[\eg][among others]{Eddington1916,Chandrasekhar1935,Chandrasekhar1960,King1956,Matsui1986,Weaver1995,Pujol2003,Chevallier2007,Shaviv2011}. However, the discovery of super-Earths, giant exoplanets, brown dwarfs and low-mass stars close to a source of intense radiation has prompted the need for solutions that account for both an outside and an inside radiation field, and properly link low and high optical depths levels. \citet{Hubeny2003}, \citet{Rutily2008}, \citet{Hansen2008}, \citet{Guillot2010}, \citet{Robinson2012} and \cite{Heng2012} provided such solutions in the framework of a semi-grey model, with one opacity for the incoming irradiation (generally mostly at visible wavelengths), and one opacity for the thermal radiation field (generally mostly at infrared wavelengths). These approximations have been used in hydrodynamical models of planetary atmospheres \citep[\eg][]{Heng2011,Rauscher2013}, planetary evolution models \citep[\eg][]{Miller-Ricci2010, Guillot2011, Budaj2012a}, planet synthesis models \citep{Mordasini2012b,Mordasini2012a}, and a variety of other applications. They  prompted \citet{Madhusudhan2009} to derive an inversion method aimed at finding all possible thermal and compositional atmospheric structures in agreement with available spectroscopic data. 

\begin{figure}[!htb]
\includegraphics[width=\linewidth]{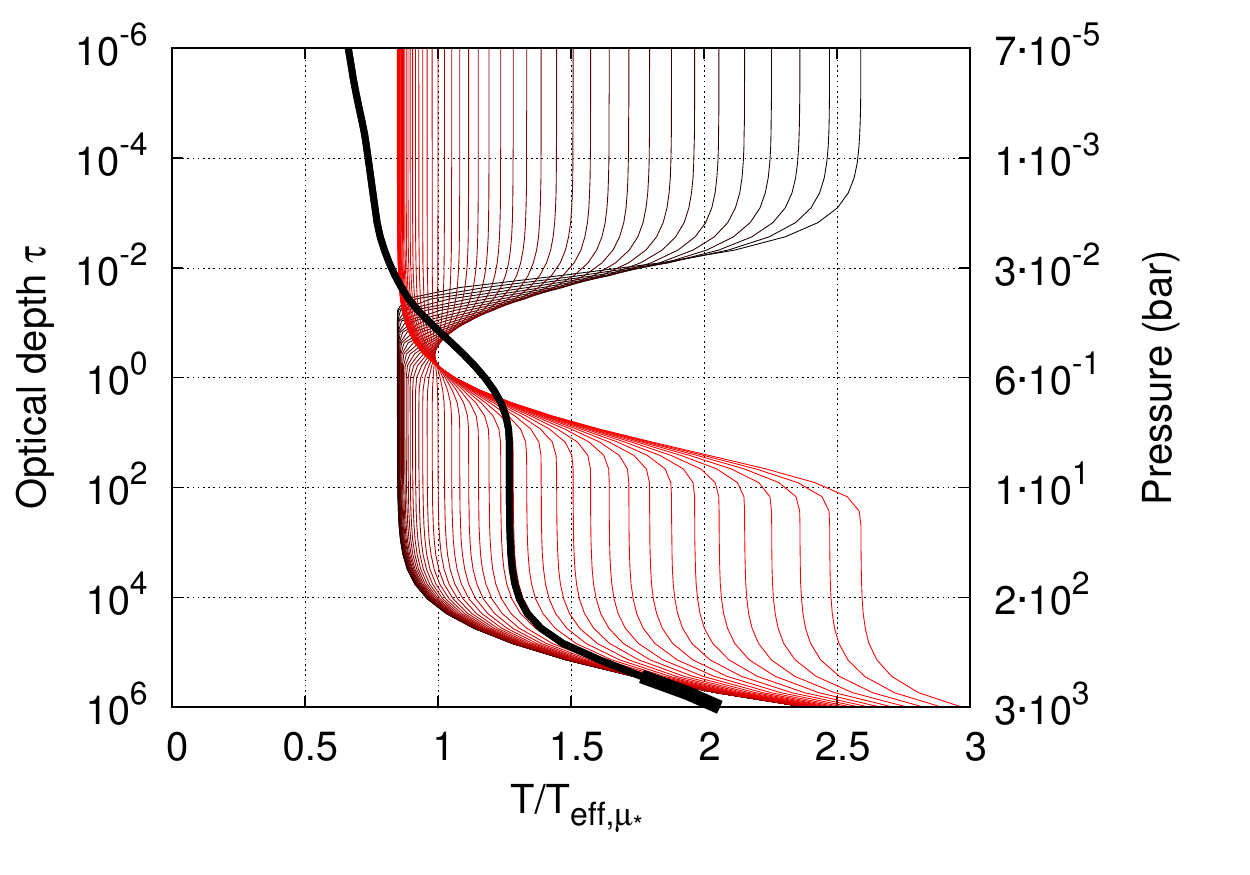}
\caption{Optical depth vs. atmospheric temperature in units of the effective temperature. A numerical solution obtained from \citet{Fortney2008} (thick black line) is compared to the semi-grey analytical solutions of \citet{Guillot2010} for values of the greenhouse factor $\Gv^{-1}$ ranging from 0.01 to 100  (black to red lines). Small values of $\Gv$ are redder. We used $\mu_{*}=1/\sqrt{3}$, $T_{\rm irr}=\unit{1250}\kelvin$ corresponding to the dayside average profile of a planet at $0.05AU$ from a sun-like star, and accounting for the albedo obtained in the numerical model. The internal temperature is $\tint=125\kelvin$ and gravity $\unit{25}\meter\rpsquare\second$. The effective temperature of the studied slice of atmosphere$^{1}$ is defined by $\teffmu^4=\tint^4+\mu_*\tirr^4$. For the numerical solution, the relation between pressure and optical depth was calculated using Rosseland mean opacities and TiO and VO opacities were not included.}
\label{fig::Profiles-G-NG}
\end{figure}

However, as shown in Fig.~\ref{fig::Profiles-G-NG} for an atmosphere irradiated from above with a flux $\sigma T_{\rm irr}^4$ and heated from below with a flux $\sigma T_{\rm int}^4$, while semi-grey models provide solutions that are well-behaved when compared to full numerical solution at optical depths larger than about unity, the temperatures at low-optical depth appear to be systematically hotter than in the numerical solutions. Most severely, this occurs {\em regardless of the choice of the two parameters of the problem}, i.e. the thermal (infrared) opacity $\kappa_{\rm th}$ and the ratio of the visible to infrared opacity $\gamma_{\rm v}\equiv \kappa_{\rm v}/\kappa_{\rm th}$. For hot Jupiters, as in the example of Fig.~\ref{fig::Profiles-G-NG}, the real temperature profiles at low optical depths can be several hundreds of Kelvins cooler than predicted by the semi-grey solutions. 

The levels probed both by transit spectroscopy and by the observations of secondary eclipses of exoplanets often correspond to low-optical depths levels \citep[\eg][]{Burrows2007,Fortney2008, Showman2009}, \ie where semi-grey models seem to systematically overestimate the temperatures. Furthermore, the fact that the problem persists regardless of the main parameters implies that the functional form of the semi-grey solutions is probably not appropriate for inversion models. Non-grey effects are known to facilitate the cooling of the upper atmosphere (see~\citet{Pierrehumbert2010} for a qualitative explanation). Obviously they must be included. This is the purpose of the present paper.

We hereafter first describe previous analytical methods used to solve the radiative transfer problem analytically. In Section~3, we then derive an analytical non-grey line model, and apply it to the structure of irradiated giant planets in Section~4. We note at this point that while we focus the discussion on exoplanets, we believe that this model is applicable to a much wider variety of problems, as long as an atmosphere is irradiated both from above and below. Out method could also be used to solve the radiative transfer equations in other geometries, such as the problem of the protoplanetary disk thermal structure. We provide our conclusions in Section~5.

\section{Assumptions and previous analytical models}

\subsection{Setting}

\subsubsection{The equation of radiative transfer}


{ Following \cite{Guillot2010}, we will consider the problem of a plane-parallel atmosphere in local thermodynamic equilibrium which receives from above a collimated flux $\sigma T_{\rm irr}^4$ at an angle $\theta_*=\cos^{-1}(\mu_{*})$ from the vertical, and from below an isotropic flux $\sigma T_{\rm int}^4$. The total energy budget of the modeled atmosphere is then set by $\teffmu^{4}=T_{\rm int}^4+\mu_{*}T_{\rm irr}^4$, which define the effective temperature in this paper\footnote{We note that in stellar physics the effective temperature is usually what we call the internal temperature. In both planetary and stellar fields, the effective temperature aim at representing the total energy budget of the atmosphere. Whereas in stellar physics most of the flux comes from the deep interior, this is no more true in irradiated atmospheres and $T_{\rm eff,\,\mu_{*}}^{4}=T_{\rm int}^{4}+\mu_{*}T_{\rm irr}^{4}$ is a better representation of the total energy budget of the studied slice of atmosphere. The energy budget of the whole atmosphere is therefore $\teff^{4}=T_{\rm int}^{4}+\tirr^{4}/4$.} }.The irradiation and intrinsic fluxes are generally characterized by very different wavelengths. Although this is not required in the solution that we propose, it is convenient to think of them as being emitted preferentially the visible and in the infrared, respectively. We will neglect scattering because it complexifies the problem dramatically and because to first order, the solution with scattering is close to the one obtained when the irradiation flux is reduced by a factor $(1-A)$, where $A$ is the Bond albedo. We note that \citet{Heng2012} provide an elegant solution to include scattering in analytical radiative transfer solutions within the two-stream approximation (see also \citet{Meador1980} for a review of the different two stream methods including scattering). We defer including this effect to a future work. 

In order to solve the radiative transfer problem for a plane parallel atmosphere in local thermodynamic equilibrium, one has to solve the following equation for all frequency $\nu$ and all directions $\mu$ \citep{Chandrasekhar1960}:
\begin{equation}
-\mu\frac{\mathrm{d}I_{\mu\nu}}{\mathrm{d}m}=\kappa_{\nu}I_{\nu\mu}-\kappa_{\nu}B_{\nu}(T)
\label{eq::RadTrans}
\end{equation}
where $I_{\mu\nu}$ is the specific intensity at the wavelength $\nu$ propagating with an angle $\theta=\cos^{-1}(\mu)$ with the vertical, $\kappa_{\nu}$ is the opacity at a given wavelength and $B_{\nu}$ is the Planck function and $\mathrm{d}m=\rho \mathrm{d}z$ is the mass increment along the path of the radiation. (As usual, $T$, $\rho$ and $z$ are the atmospheric temperature, density and height, respectively.)
The main difficulty in solving eq.~\eqref{eq::RadTrans} lies in its triple dependence on $\mu$, $\nu$ and $T$ and additional dependence on $m$. An analytical solution requires simplifications in terms both of the opacities used and/or dependence of the radiation intensity with angle. 

\subsubsection{Opacities and optical depth}
\label{sec::Opacities}

The need for simplification implies that means of the opacities must be used. The first one by importance is the Rosseland mean, defined as:
\begin{equation}
\frac{1}{\Kr}\equiv\left(\int_{0}^{\infty} \frac{\partial B_{\nu}}{\partial T}\,\mathrm{d}\nu\right)^{-1}\int_{0}^{\infty}\frac{1}{\Knu}\frac{\partial B_{\nu}}{\partial T}\,\mathrm{d}\nu
\label{eq::Rosseland}
\end{equation}
It can be demonstrated that when at all wavelengths the mean free path of photons is small compared to the scale height of the atmosphere, the radiative gradient obeys its well-defined diffusion limit and (unless convection sets in) the temperature profiles becomes that obtained from a grey atmosphere in which the opacity is set to the Rosseland mean \citep[][p. 350]{Mihalas1984}. 
We hence define the optical depth $\tau$ on the basis of the Rosseland mean opacity, such that, along the vertical direction:
\begin{equation}
\mathrm{d}\tau\equiv\Kr\mathrm{d}m
\label{eq::dtau}
\end{equation}
Assuming hydrostatic equilibrium, the relation between pressure and optical depth can be found by integrating equation  \eqref{eq::dtau}:
\begin{equation}
\tau(P)=\int_{0}^{P}\frac{\Kr(P',T(P'))}{g}\D P'
\label{eq::tauP}
\end{equation}
The optical depth thus becomes the natural variable to account for the dependence with depth in the radiative transfer problem. For any strictly positive Rosseland mean opacities, equation \eqref{eq::tauP} is a bijection relating pressure and optical depth. Thus, any solution of the radiative transfer equations in terms of optical depth can be converted to a solution in term of pressure for any functional form of the Rosseland mean opacities.

The second mean of the opacity that is traditionally used for radiative transfer is the so-called Planck mean:
\begin{equation}
\Kp\equiv\left(\int_{0}^{\infty} B_{\nu}\,\mathrm{d}\nu\right)^{-1}\int_{0}^{\infty}\kappa_{\nu}B_{\nu}\,\mathrm{d}\nu
\label{eq::Planck}
\end{equation}
We use it to quantify the ``non-greyness'' of the atmosphere through the ratio between the Planck and Rosseland means:
\begin{equation}
\Gp\equiv\frac{\Kp}{\Kr}
\label{eq:Gp}
\end{equation}
While the value of the Rosseland mean opacity is dominated by the smallest values of the opacity function $\kappa_{\nu}$, the Planck mean opacity is dominated by its highest values. Thus, it can be shown that $\Gp=1$ for a grey atmosphere and $\Gp>1$ for a non-grey atmosphere \citep[][]{King1956}. 

In irradiated atmospheres, a collimated flux coming from the star is absorbed at different atmospheric levels. We name $\kappa_{\rm v}$ the opacity relevant to the absorption of the stellar flux. As will be shown in Sec.~\ref{sec::MultipleBands}, the absorption of the visible flux appear linearly in the radiative transfer equations. Thus a solution can be found using multiple visible opacity bands $\kappa_{\rm v1}$, $\kappa_{\rm v2}$, etc.

We further define the ratio of the visible opacity to the mean (Rosseland) thermal opacity: 
\begin{equation}
\Gv\equiv\Kv/\Kr.
\label{eq:Gv}
\end{equation}
In order to solve the radiative transfer problem analytically, we suppose that $\Gv$ is constant with optical depth. Once $\Gv$ is chosen, we can solve the equations for the visible radiation independently from the final thermal structure of the atmosphere. Of course, purely grey models are such that $\Gv=1$.

\subsubsection{The picket-fence model}
\label{sec::LineModel}

\begin{figure}[!htb]
\includegraphics[width=\linewidth]{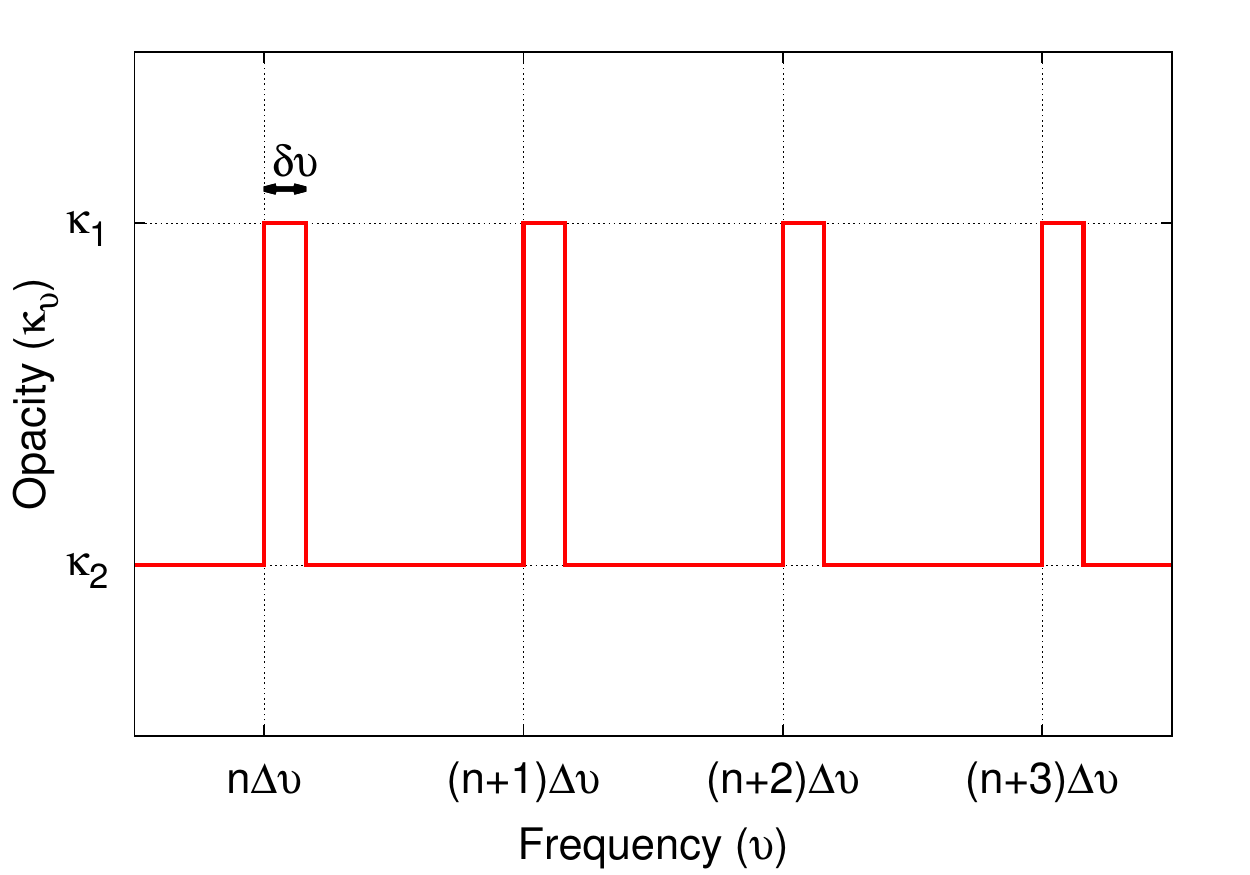}
\caption{Simplified thermal opacities for the  picket-fence model. $\beta=\delta\nu/\Delta\nu$ is the equivalent bandwidth (see text).}
\label{fig::Opa-schema}
\end{figure}

It is important to note at this point that two set of opacities with different wavelength-dependence may have the same Rosseland {\em and\/} Planck means. We must constrain the problem further, and to this intent, we now consider the simplest possible line model, {known as the picket-fence model~\citep{Mihalas1978}}, where the thermal opacities can take two different values $\Ka$ and $\Kb$ (see Fig.\ref{fig::Opa-schema}) such that:
\begin{equation}
\Knu= \left\{ 
\begin{array}{l l}
  \Ka & \quad \mbox{for $\nu\in[n\Delta\nu,n\Delta\nu+\delta\nu[$}\\
  \Kb & \quad \mbox{for $\nu\in[n\Delta\nu+\delta\nu,(n+1)\Delta\nu[$}\\ 
\end{array} \right. \hspace{0.3cm} n\in[1,N]
\label{eq::CombOpacities}
\end{equation}
We define an equivalent bandwidth by:
\begin{equation}
\beta=\frac{1}{\int_{0}^{\infty} B_{\nu}\,\mathrm{d}\nu}\sum_{n=1}^{N}\int_{n\Delta\nu}^{n\Delta\nu+\delta\nu}B_{\nu} \,\mathrm{d}\nu
\label{eq::beta}
\end{equation}
The characteristic width of the Planck function can be defined as $\Delta\nu_{P}\equiv \left(\frac{1}{B}\frac{\partial B_{\nu}}{\partial\nu}\right)^{-1}$. When choosing $\Delta\nu<<\Delta\nu_{P}$, the Planck function can be considered constant over $\Delta\nu$ and we get $\beta=\delta\nu/\Delta\nu$. The Planck and Rosseland mean opacities then become (see eqs.~\eqref{eq::Rosseland} and \eqref{eq::Planck}):
\begin{equation}
\Kr=\frac{\Ka\Kb}{\beta\Kb+(1-\beta)\Ka}
\end{equation}
\begin{equation}
\Kp=\beta\Ka+(1-\beta)\Kb
\end{equation}
We also define the following ratios:
\begin{equation}
\Ga\equiv{\Ka}/{\Kr}
\end{equation}
\begin{equation}
\Gb\equiv{\Kb}/{\Kr}
\end{equation}
\begin{equation}
R\equiv{\Ka}/{\Kb}={\Ga}/{\Gb}
\end{equation}
Following \cite{Chandrasekhar1935}, we also define a limit optical depth 
\begin{equation}
\taulim\equiv\frac{1}{\Ga\Gb}\sqrt{\frac{\Gp}{3}}
\label{eq::taulim}
\end{equation}


\subsection{The method of discrete ordinates for the non-irradiated problem}

\subsubsection{The grey case}
An approximate method to solve eq.~(\ref{eq::RadTrans}) including the angular dependency has been developed by \citet{Chandrasekhar1960} in the case of a non-irradiated atmosphere ($\tirr=0$). The idea is to replace the integrals over angle in eq.~(\ref{eq::RadTrans}) by a gaussian sum over $\mu$. It can then be solved to an arbitrary precision by increasing the number of terms in the sum. The boundary condition at the top of the atmosphere is simply given as $I_{\mu<0}(0)=0$. The expansion to the fourth term yields the following temperature profile:
\begin{equation}
T(\tau)^{4}=\frac{3\tint^{4}}{4}\left(\tau+Q+L_{1}e^{-k_{1}\tau}+L_{2}e^{-k_{2}\tau}+L_{3}e^{-k_{3}\tau}\right)
\end{equation}
with $Q=0.706920$, $L_{1}=-0.083921$, $L_{2}=-0.036187$, $L_{3}=-0.009461$, $k_1=1.103188$, $k_2=1.591778$ and $k_3=4.45808$ \citep[][ table VIII]{Chandrasekhar1960}\footnote{We noticed that the values of $L_{1}$ and $L_{3}$ in \citeauthor{Chandrasekhar1960}'s book were inverted and corrected this here.}. 
One of the strong results from this formalism is that the skin temperature of the planet, the temperature at zero optical depth, is independent of the order of expansion and therefore corresponds to the exact value:
\begin{equation}
\tskin^{4}=\sqrt{3}\frac{\tint^{4}}{4}
\label{eq::SkinTempGreyExact}
\end{equation}
This expression is exact only in the limit of a grey, non-irradiated atmosphere. 

\subsubsection{The non-grey case}
\citet{Chandrasekhar1960} further developed a perturbation method in order to include non-grey thermal opacities. This method was improved by \citet{Krook1963}. However, these perturbation methods either work for small departure from the grey opacities or involve a fastidious iterative procedure \citep[\eg][]{Unno1960,Avrett1963} and are no longer fully analytical. However, considering that the variations in the opacities are short compared to the variations of the Planck function, analytical solutions can be found for an arbitrary large departure from the grey opacities. Noting the similar role of $\mu$ and $\kappa_{\nu}$ in eq.~(\ref{eq::RadTrans}), \citet{King1956}, following \cite{Munch1946} used the method of discrete ordinates in order to turn the integrals over frequency into gaussian sums. For the picket-fence model defined in Section~\ref{sec::LineModel} and the second approximation for the angular dependency, King's method leads to the following temperature profile:
\begin{equation}
T^{4}(\tau)=\frac{3}{4}\tint^{4}\left[\frac{1}{\sqrt{3\Gp}}+\tau+\frac{(\sqrt{\Gp}-\Ga)(\sqrt{\Gp}-\Gb)}{\Ga\Gb\sqrt{3\Gp}}(e^{-\tau/\taulim}-1)\right]
\label{eq::T4King}
\end{equation}
As in the grey case, the method of discrete ordinates leads to an exact relation for the skin temperature, whatever the dependency of $\kappa_{\nu}$ with frequency (but no dependence in pressure or temperature):
\begin{equation}
\tskin^{4}=\sqrt{\frac{3}{\Gp}}\frac{\tint^{4}}{4}
\label{eq::TskinKing}
\end{equation}
In the grey limit, $\Gp=1$ and we recover eq.~\eqref{eq::SkinTempGreyExact}. Otherwise, $\Gp>1$, implying that for a non-irradiated atmosphere, non-grey effects will always tend to lower the atmospheric skin temperature.

\subsection{Moment equation method}

\subsubsection{Equations for the momentum of the radiation intensity}


\label{sec::Eddington}
A simpler way to solve the radiative transfer equation has been carried out by \citet{Eddington1916}. The idea is to solve the equation using the different momentum of the intensity defined as:
\begin{equation}
(J_{\nu}, H_{\nu}, K_{\nu})=\int_{-1}^{1}I_{\mu\nu}(1,\mu,\mu^{2})\,\D\mu
\label{eq:MomentDef}
\end{equation}
Then, integrating over $\nu$ and $\mu$ eq.~\eqref{eq::RadTrans} and $\mu$ times eq.~\eqref{eq::RadTrans} one gets the momentum equations:
\begin{equation}
\frac{\D H_{\nu}}{\D\tau_{\nu}}=J_{\nu}-B_{\nu}(T)
\label{eq::dJnu}
\end{equation}
\begin{equation}
\frac{\D K_{\nu}}{\D\tau_{\nu}}=H_{\nu}
\label{eq::dKnu}
\end{equation}
Assuming the atmosphere to be in radiative equilibrium, we can write:
\begin{equation}
\int_{0}^{\infty}(\kappa_{\nu}J_{\nu}-\kappa_{\nu}B_{\nu})\,\mathrm{d}\nu=0
\label{eq::RadEq}
\end{equation}
For a grey atmosphere ($\kappa_{\nu}=\Kr\ \forall\nu$), eqs.~(\ref{eq::dJnu}--\ref{eq::RadEq}) can be integrated over frequency, leading to an equation on $J$, $H$, $K$ and $B$, the frequency-integrated version of $J_{\nu}$, $H_{\nu}$, $K_{\nu}$ and $B_{\nu}$. The radiative equilibrium equation becomes:
\begin{equation}
J=B
\label{eq::RadEqGrey}
\end{equation}

Equations~(\ref{eq::dJnu}--\ref{eq::RadEq}) are then a set of three equations with four unknowns. The system is not closed because by integrating eq.~(\ref{eq::RadTrans}) over all angles we have lost the information on the angular dependency of the irradiation. A closure relationship that contains angular dependency of the radiation field is therefore needed. A common closure relationship, known as the Eddington approximation is:
\begin{equation}
J=3K
\end{equation}
This relationship is exact in two very different cases: when the radiation field is isotropic ($I_{\mu}$ independent of $\mu$) and in the two-stream approximation ($I_{\mu>0}=I_{0}^{+}$ and $I_{\mu<0}=I_{0}^{-}$)
Although this seems a very restrictive approximation, it is relevant for the deep layers of the atmosphere due to the quasi-isotropy of the radiation field there. It is also good for the top of the atmosphere, where the flux comes mainly from the $\tau\approx 1$ layer. Indeed, the exact solution gives a ratio $J/K$ that differs by no more than $20\%$ from the $1/3$ ratio over the whole atmosphere and leads to a temperature profile which is correct to $4\%$ in the grey case (see the plain blue line of Fig. \ref{fig::Chand} hereafter).

\subsubsection{Top boundary condition}
\label{sec::BoundaryCondition}
Although in the method of discrete ordinates the boundary condition at the top of the atmosphere is intuitive, in the momentum equations method it is less obvious and different choices has been made by different authors. Usually, the expression for $J$ is known and some integration constant needs to be found. Two equations are needed, one for $J(0)$ and one for $H(0)$. Four possibilities are widely used in the literature from which one has to choose two:
\begin{enumerate}
\item The radiative equilibrium equation that relates the emergent flux at the top of the atmosphere to the internal flux from the planet and the incident flux from the star.
\item An ad-hoc relation between $H(0)$ and $J(0)$ at $\tau=0$: $H(0)=f_{\rm H}J(0)$, where $f_{\rm H}$ is often called the second Eddington coefficient. 
\item A calculation of $H(0)$ from the second moment equation (eq. \eqref{eq::dKnu}) and the Eddington approximation.
\item A calculation of $H(0)$ from the integration of the source function through the entire atmosphere, known as the Milne equation \citep[][p. 347]{Mihalas1984}: $ 
H(0)=\frac{1}{2}\int_{0}^{\infty}B\left(\tau\right)E_{2}(\tau)\,\mathrm{d}\tau $
\end{enumerate}

For grey and semi-grey models, the first condition is natural. Therefore it is used by \citet{Hansen2008} and \citet{Guillot2010}. For the other part of the top boundary condition, \citet{Guillot2010} choose to use the second and \citet{Hansen2008} the fourth condition. (See Appendix A of \citet{Guillot2010} for a comparison of the expressions). 

In the case of a non-grey model, the first condition cannot be implemented (at least directly) because it is a constraint on the total thermal flux but it provides no information on how the thermal flux is split between the opacity bands that are considered. \citet{Chandrasekhar1935} therefore uses for his non-grey, non-irradiated model conditions 2 and 3 in each of the opacity band. He further notes that using condition 4 instead of condition 3 should yield better results, but leads to more complex expressions. In this work, because an accurate treatment of the flux is needed for the non-grey irradiated model, we will use conditions 2 and 4 in each of the opacity band. All these models are discussed in the next sections and summarized in table~\ref{table::Models}.

\subsubsection{Non-irradiated, grey case}
In this section we consider the case $\tirr=0$. Under the grey approximation, using the conditions~1 and 4, it can be shown that \citep[][p.357]{Mihalas1984}:
\begin{equation}
T^{4}(\tau)=\frac{3}{4}\tint^{4}\left(\frac{2}{3}+\tau\right)
\end{equation}
which leads to the same solution than assuming condition~2 with $\fh=1/2$. The skin temperature is then:
\begin{equation}
\tskin^{4}=2\frac{\tint^{4}}{4}
\label{eq::TskinEdd}
\end{equation}
which differs from the exact solution [eq.~(\ref{eq::SkinTempGreyExact})] by a factor ${\sqrt{3}}/{2}$. 
Assuming $\fh=1/\sqrt{3}$ is thus tempting, as it leads to the correct skin temperature, but, on the other hand, it leads to a temperature profile which is less accurate around $\tau\approx1$.

\subsubsection{Non-irradiated non-grey {picket-fence model}}
\citet{Chandrasekhar1935} provides solutions to the moment equations for the picket-fence model presented in Section~\ref{sec::LineModel}. He assumes that the relation $H(0)=\frac{1}{\fh}J(0)$ with $\fh=1/2$, valid in the grey case under the Eddington approximation [see eq.~(\ref{eq::TskinEdd})], holds for the two thermal channels separately. Using this condition together with condition~3 he obtains the following temperature profile:
\begin{equation}
\begin{split}
T^{4}(\tau)&=\frac{3\tint^{4}}{4}\left[\tau+\frac{\frac{2}{3}+\sqrt{\frac{1}{3\Gp}}}{1+\frac{1}{2}\sqrt{3\Gp}}\right]\\
	&+\frac{3\tint^{4}}{4}\left(\frac{\Gp-1}{\sqrt{\Gp}}\right)\frac{\frac{1}{\sqrt{3}}+\sqrt{\Gp}\taulim}{1+\frac{1}{2}\sqrt{3\Gp}}\left(1-e^{-\tau/\taulim}\right)
\end{split}
\label{eq::T4Chandra}
\end{equation}
and the equation for the skin temperature:
\begin{equation}
T_{\rm skin}^{4}=2\left(\frac{2+\sqrt{\frac{3}{\Gp}}}{2+\sqrt{3\Gp}}\right)\frac{T_{\rm int}^{4}}{4}
\label{eq::TskinChandra}
\end{equation}
As expected, this equation reduces to eq.~(\ref{eq::TskinEdd}) in the limit $\Gp=1$. 
For large values of $\Gp$, it is easy to show that this relation differs by a factor $4/3$ from the exact one derived with the method of discrete ordinates. As for the grey case, using $\fh=1/\sqrt{3}$ would lead to the exact solution for the skin temperature, but at the expense of the accuracy of the profile at deeper levels.  Again, we note that, in the non-irradiated case, the temperature at the top of the atmosphere is determined by a single parameter, $\Gp$, representing the ``non-greyness'' of the atmosphere.

\subsubsection{Irradiated semi-grey model}

In the case of irradiated atmospheres, the presence of an incoming collimated flux at the top of the atmosphere breaks the angular symmetry of the equations. The radiative transfer problem thus cannot be solved analytically (at least not in a simple way) through the discrete ordinates technique any longer. The momentum method is thus required. 

In order to solve the problem, the radiation field is split into two parts: The incoming, collimated radiation field on one hand, the thermal radiation field on the other. The radiative equilibrium equation (eq. \eqref{eq::RadEq}) links the two streams as can be seen in Section~\ref{sec::AnalyticalModel} \citep[see also][]{Hansen2008,Guillot2010,Robinson2012}. As mentioned previously in Section~\ref{sec::Opacities}, when the incident radiation is at much shorter wavelength than the thermal emission of the atmosphere, the two streams correspond to different characteristic wavelengths and may often be labelled as ``visible'' and ``infrared''. This not a requirement however: the solutions apply if the radiation field correspond to other wavelengths or if they overlap. 

As discussed previously (Section~\ref{sec::BoundaryCondition}), the boundary condition at the top of the model can be chosen in several ways. When using condition 2, \cite{Guillot2010} lets the value of $\fh$ be either $1/2$ or $1/\sqrt{3}$, based on the values obtained in the non-irradiated case. $\fh=1/2$ is the value that arises from the calculation of the angle dependence between $H(0)$ and $J(0)$ in the isotropic case, but $\fh=1/\sqrt{3}$ provides a skin temperature that agrees with the exact value. The two solutions differ by $\approx3\%$ at most (see Fig.~\ref{fig::Guillot}), and choosing one over another is not crucial.
In any case, for an easier comparison, we provide here the solution of \citet{Guillot2010} for $\fh=1/2$:
\begin{equation}
\begin{split}
T^4=&{3\tint^4\over 4}\left[{2\over 3}+\tau\right]\\
&+{3\tirr^4\over 4}\mu_*\left[{2\over 3}+  {\mu_*\over \Gv}+\left({\Gv\over 3\mu_*}-{\mu_*\over\Gv}\right) e^{-\Gv\tau/\mu_*}\right].
\label{eq::T4Guillot}
\end{split}
\end{equation}
where $\mu_{*}$ is the cosine of the angle of the incident radiation. The skin temperature is:
\begin{equation}
T_{\rm skin}^{4}=2 \frac{\tint^{4}+\mu_{*}\tirr^{4}}{4}+\Gv\tirr^{4}
\label{eq::Tskin-Guillot}
\end{equation}
For $\Gv\to 0$, the incident radiation is absorbed in the deep layers of the atmosphere and the skin temperature converges to the skin temperature of a grey model with an effective temperature $\teffmu^{4}=\tint^{4}+\mu_{*}\tirr^{4}$. The semi-grey model depends only on the parameter $\Gv$. 

\begin{figure}[!htb]
\includegraphics[width=\linewidth]{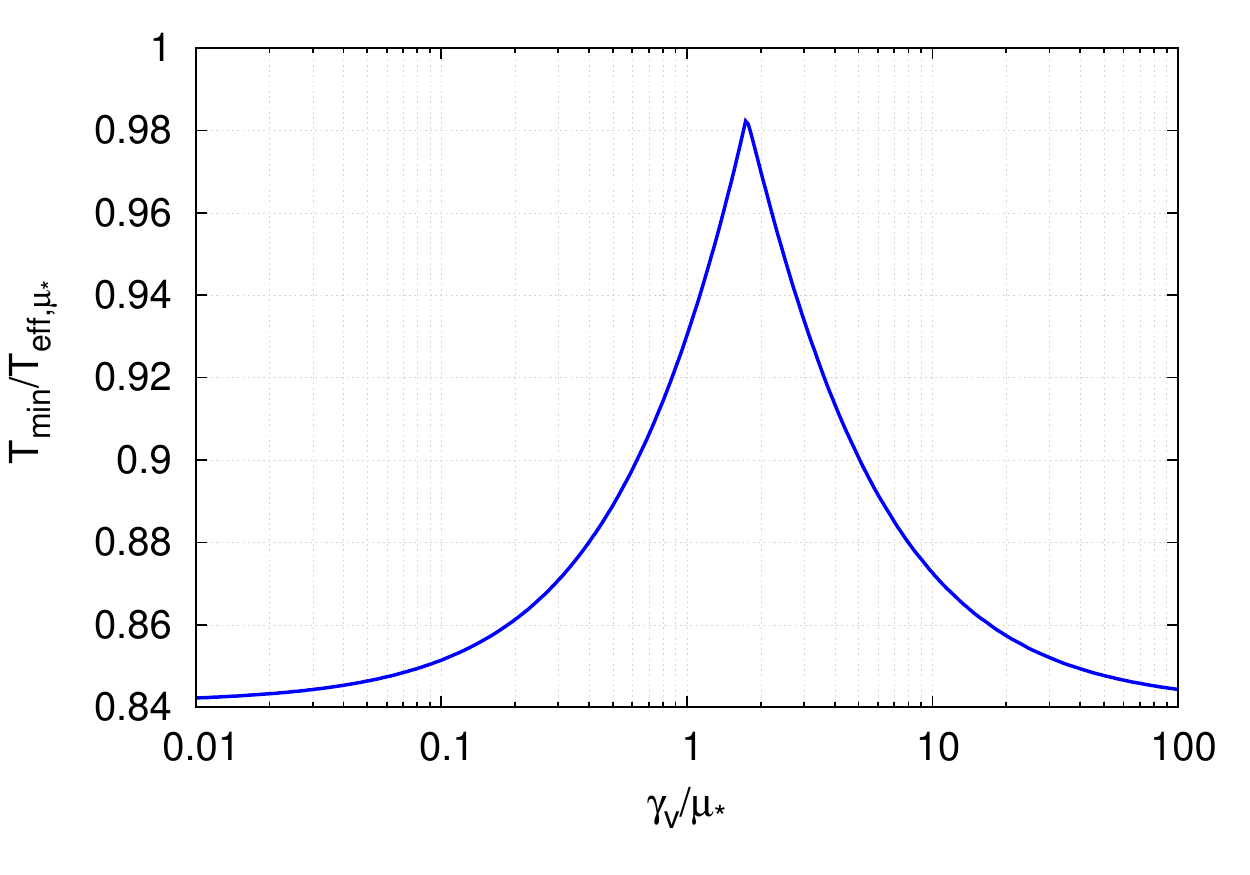}
\caption{Minimum temperature of the semi-grey model in terms of the effective temperature as a function of $\Gv/\mu_{*}$.}
\label{fig::Grey}
\end{figure}

As discussed in introduction (see Fig.~\ref{fig::Profiles-G-NG}), the semi-grey model predicts minimum temperatures that are generally higher than numerical solutions for irradiated exoplanets, independently of the choice of $\Gv$. 
In fact, similarly to the skin temperature, the minimum temperature of a semi-grey atmosphere, shown in Fig.~\ref{fig::Grey}, depends only on the values of $\teffmu$ and $\Gv$. It is lowest and equal to $\teffmu/2^{1/4}$ both in the $\Gv\to 0$ and $\Gv\to \infty$ limits. This lower bound for the semi-grey temperature profile is hotter than what the one obtained by numerical calculations taking into account the full set of opacities. The discrepancy is much larger than the variations resulting from the approximation of the momentum method. Clearly, non-grey effects must be invoked to explain the low temperatures obtained by numerical models at low optical depths. 

\begin{center}
\begin{table*}

\label{table::Models}
\caption{Summary of the different models compared in this paper.}
\centering
\ra{1.2} 
\begin{tabular}{>{\flushleft}m{4cm} >{\centering} m{1.4cm} >{\centering} m{1.4cm} >{\centering} m{1.4cm} >{\centering} m{3cm} >{\centering} m{2cm}}

\toprule 
Model & External irradiation  & Eddington approx. & Non-grey thermal opacities &Top boundary condition & Expression  \\ 
\midrule
Numerical &$\surd$ & $\times$  & $\surd$ & $I(0)=I_{\rm star}$ for $\mu<0$ & N/A\\
\citet{King1955} &$\times$ & $\times$ & $\surd$ & $I(0)=0$ for $\mu<0$  & eq.~(\ref{eq::T4King}) \\
\citet{Chandrasekhar1935}&$\times$ & $\surd$ & $\surd$ & eq (\ref{eq::dKnu}) \& $f_{\rm H}=1/2$ & eq.~(\ref{eq::T4Chandra})  \\
\cite{Hansen2008} &$\surd$ & $\surd$ & $\times$ & Rad. eq. \& eq. (\ref{eq::H12})  & ---  \\
\citet{Guillot2010}, $\fh=1/2$ &$\surd$ & $\surd$ & $\times$ & Rad. eq. \& $\fh=1/2$  & eq.~(\ref{eq::T4Guillot})  \\
\citet{Guillot2010}, $\fh=1/\sqrt{3}$ &$\surd$ & $\surd$  & $\times$ & Rad. eq. \& $\fh=1/\sqrt{3}$  & ---  \\
This model &$\surd$ & $\surd$  & $\surd$ & eq. (\ref{eq::H12}) \& $\fh=1/2$  & eq.~(\ref{eq::Tprofile})  \\
\bottomrule 

\end{tabular}
 
 \end{table*}

\end{center}
\section{An analytical irradiated non-grey picket-fence model}
\subsection{Equations}
\label{sec::AnalyticalModel}

We now derive the equations for an irradiated atmosphere in local thermodynamic equilibrium with infrared line opacities as described in sec. \ref{sec::Opacities}. Thus, our model contains three different opacities: $\Ka$ and $\Kb$ for the thermal radiation and $\Kv$ relevant for the incoming radiation of the star. As explained before, the difference between the thermal and the visible channel is based on the angular dependency of the radiation and not on the frequency. Although the method of discrete ordinates is shown to lead to more exact results, it is complex to adapt to the irradiated case. Therefore, following \cite{Chandrasekhar1935} and \cite{Guillot2010}, we solve the radiative transfer equations using the momentum equations.
Integrating eqs.~(\ref{eq::dJnu}) and (\ref{eq::dKnu}) over each thermal band we obtain:

\begin{equation}
\frac{\D H_{\rm (1,2)}}{\D\tau}=\gamma_{\rm (1,2)}J_{\rm (1,2)}-\gamma_{\rm (1,2)}(\beta,1-\beta)B(T)
\label{eq::dJnu12}
\end{equation}

\begin{equation}
\frac{\D K_{\rm (1,2)}}{\D\tau}=\gamma_{\rm (1,2)}H_{\rm (1,2)}
\label{eq::dKnu12}
\end{equation}

Where the subscript indicates the integrated quantities over the given thermal band. Thus for a quantity $X_{\nu}$ we have: 
\begin{equation}
X_{\rm 1}=\sum_{n}\int_{n\Delta\nu}^{n\Delta\nu+\delta\nu}X_{\nu} \,\mathrm{d}\nu
\end{equation}

\begin{equation}
X_{\rm 2}=\sum_{n}\int_{n\Delta\nu+\delta_{\nu}}^{(n+1)\Delta\nu}X_{\nu} \,\mathrm{d}\nu
\end{equation}

The Planck function is considered constant over each bin of frequency $\Delta\nu$ and therefore $B_{1}=\beta B$ and $B_{2}=(1-\beta)B$. 
We now assume that the Eddington approximation is valid in the two bands separately: $J_{\rm (1,2)}=3K_{\rm (1,2)}$. Equations \eqref{eq::dJnu12} and \eqref{eq::dKnu12} can be combined into:
\begin{equation}
\frac{d^{2}\Ja}{d\tau^{2}}=3\Ga^{2}(\Ja-\beta B)
\label{eq::J1}
\end{equation}
\begin{equation}
\frac{d^{2}\Jb}{d\tau^{2}}=3\Gb^{2}(\Jb-(1-\beta) B)
\label{eq::J2}
\end{equation}
and the radiative equilibrium equation becomes:
\begin{equation}
\Ga\Ja+\Gb\Jb+\Gv\Jv=\Gp B,
\label{eq::radtran}
\end{equation}
where the quantities with subscript $\rm v$ are the momentum of the incident stellar radiation. They can be directly calculated when assuming that the incoming stellar radiation arrives as a collimated flux and hit the top of the atmosphere with an angle $\theta_{*}=\cos^{-1}\mu_{*}$:
\begin{equation}
(J_{\rm v}, H_{\rm v}, K_{\rm v})=(1,\mu_{*},\mu_{*}^{2})I_{*}
\end{equation}
and $I_{*}=\int_{0}^{\infty}I_{*\nu}\,\D\nu$ is the total incident intensity. 

The absorption of the stellar irradiation can be treated separately from the thermal radiation and $\Jv$ is given by eq.~(13) of \citet{Guillot2010}: 
\begin{equation}
\Jv(\tau)=-\frac{H_{\rm v}(0)}{\mu_{*}}e^{-\Gv^*\tau}
\label{eq::Jv}
\end{equation}
where we have simplified the notation by introducing the parameter $\Gv^{*}\equiv\Gv/\mu_{*}$. 

Equations (\ref{eq::J1}) to (\ref{eq::radtran}) are a set of three coupled equations with three unknowns $\Ja$, $\Jb$, $B$. In order to decouple these equations we define two new variables:
\begin{equation}
\begin{spreadlines}{12pt}
\left\{
\begin{aligned} 
\JG\equiv\frac{\Ja}{\Ga}+\frac{\Jb}{\Gb}\\
\JGb\equiv\frac{\Ja}{\Ga^{3}}+\frac{\Jb}{\Gb^{3}}
\label{sys::JGab}
\end{aligned}
\right.
\end{spreadlines}
\end{equation}
Conversely we can come back to the original variables:
\begin{equation}
\left\{
\begin{spreadlines}{12pt}
\begin{aligned}
\Ja=-\frac{\Ga^{3}(\Gb^{2}\JGb-\JG)}{\Ga^{2}-\Gb^{2}}\\
\Jb=\frac{\Gb^{3}(\Ga^{2}\JGb-\JG)}{\Ga^{2}-\Gb^{2}}
\label{eq::JG2J}
\end{aligned}
\end{spreadlines}
\right.
\end{equation}
Using the combination of equations $\frac{1}{\Ga}$(\ref{eq::J1})+$\frac{1}{\Gb}$(\ref{eq::J2}) and equation \eqref{eq::radtran} we get:
\begin{equation}
\frac{d^{2}\JG}{d\tau^{2}}=-3\Gv\Jv
\label{eq::JGa}
\end{equation}

The combination of equations $\frac{1}{\Ga^{3}}$(\ref{eq::J1})+$\frac{1}{\Gb^{3}}$(\ref{eq::J2}) yields:
\begin{equation}
\frac{d^{2}\JGb}{d\tau^{2}}=3(\JG-B)
\label{eq::JGb}
\end{equation}
Noting that $\Ga\Ja+\Gb\Jb=(\Ga^{2}+\Gb^{2})\JG-(\Ga\Gb)^{2}\JGb$, eq.~(\ref{eq::radtran}) becomes:
\begin{equation}
B=\frac{\Ga^{2}+\Gb^{2}}{\Gp}\JG-\frac{(\Ga\Gb)^{2}}{\Gp}\JGb+\frac{\Gv}{\Gp}\Jv
\label{eq::radeqJG}
\end{equation}
Equations (\ref{eq::JGa}), (\ref{eq::JGb}) and (\ref{eq::radeqJG}) are now a set of two uncoupled differential equations and a linear equation.

\subsection{Boundary conditions}
In order the solve the differential equations we need to specify the boundary conditions. 
When $\tau\rightarrow +\infty$ we want to fulfill the diffusion approximation: $J_{\nu}=B_{\nu}$ \citep[][p. 350]{Mihalas1984}. In our case this translates to $\Ja=\beta B$ and $\Jb=(1-\beta)B$. Furthermore, at these levels, the gradient of B should also obey the diffusion approximation \citep{Mihalas1984}:
\begin{equation}
\frac{dB}{d\tau}\underset{\tau\to+\infty}\sim3H_{\infty}
\label{eq::BC1}
\end{equation}
where $4\pi H_{\infty}=\sigma \tint^{4}$ is the thermal flux coming from the interior of the planet. Using the system of equations (\ref{sys::JGab}) and noting that $\frac{\beta}{\Ga}+\frac{1-\beta}{\Gb}=1$, we can derive a condition on $\JG$ and $\JGb$:
\begin{equation}
 \frac{d\JG}{d\tau}\underset{\tau\to+\infty}\sim3H_{\infty}
 \label{eq::BC1JG}
\end{equation}
\begin{equation}
\frac{d\JGb}{d\tau}\underset{\tau\to+\infty}\sim\left(\frac{\beta}{\Ga^{3}}+\frac{1-\beta}{\Gb^{3}}\right)3H_{\infty}
 \label{eq::BC1JGb}
 \end{equation}
 For $\tau \rightarrow 0$ we specify the geometry of the intensity by setting:
 \begin{equation}
J_{\rm (1,2)}(0)=2H_{\rm (1,2)}(0)
\label{eq::BC2}
\end{equation}
Furthermore, we calculate the flux at the top of the atmosphere in each band using equation~(79.21) from \citet{Mihalas1984}. From the assumption of local thermodynamic equilibrium, the source function in the two bands is $S_{\rm 1}(\tau_{\rm 1})=\beta B(\tau/\Ga)$ and $S_{\rm 2}(\tau_{\rm 2})=(1-\beta)B(\tau/\Gb)$. The upper boundary condition on the flux of the 2 bands thus becomes:
\begin{equation}
H_{(\rm 1,\rm 2)}(0)=\frac{1}{2}\int_{0}^{\infty}(\beta,(1-\beta))B\left(\frac{\tau}{\Gab}\right)E_{2}(\tau)\,\mathrm{d}\tau
\label{eq::H12}
\end{equation}

\subsection{Solution}
The solution of a second order differential equation with constant coefficient is the sum of the solutions of the homogeneous equation and a particular solution of the complete equation. Thus, solutions of eq.~(\ref{eq::JGa}) must be of the form:
\begin{equation}
\JG(\tau)=\Ca+\Cb\tau+\frac{3}{\Gv^*}H_{\rm v}(0)e^{-\Gv^*\tau}
\label{eq::SolJGa}
\end{equation}
applying the boundary condition eq.~(\ref{eq::BC1JG}), we get $\Cb=3H$. For $\tau=0$ we obtain:
\begin{equation}
\JG(0)=\Ca+\frac{3}{\Gv^*}\Hv(0)
\label{eq::BCJG}
\end{equation}
Using equation \eqref{eq::radeqJG} to eliminate $B$ and replacing $\JG$ by its solution, eq.~(\ref{eq::JGb}) becomes:
\begin{equation}
\begin{split}
\frac{d^{2}\JGb}{d\tau^{2}}-3\frac{(\Ga\Gb)^{2}}{\Gp}\JGb&=3\left(1-\frac{\Ga^{2}+\Gb^{2}}{\Gp}\right)(\Ca+3H\tau)\\
 &+3\left(1-\frac{\Ga^{2}+\Gb^{2}}{\Gp}\right)\frac{3}{\Gv^*}\Hv(0)e^{-\Gv^*\tau}\\
 &+3\frac{\Gv^*}{\Gp}\Hv(0)e^{-\Gv^*\tau} 
\end{split}
\label{eq::DiffJGtot}
\end{equation}
Again, solutions of this differential equation must be the sum of the solutions of the homogeneous equation and one solution of the complete equation.
The homogeneous solution must have the form:
\begin{equation}
J_{\rm\gamma^{3}H}=\Cc e^{-\tau/\taulim}+\Cd e^{+\tau/\taulim}
\end{equation}
where $\Cc$ and $\Cd$ are constants of integration to be determined using the boundary conditions.
We look for a particular solution formed by the superposition of an exponential and an affine function. The affine function must then be a solution of eq.~(\ref{eq::DiffJGtot}) with $\Hv(0)=0$:
\begin{equation}
J_{\rm \gamma^{3} P1}=-\frac{\Gp}{(\Ga\Gb)^{2}}\left(1-\frac{\Ga^{2}+\Gb^{2}}{\Gp}\right)(\Ca+3H\tau)
\end{equation}
and the exponential function must be solution of eq.~(\ref{eq::DiffJGtot}), keeping only the exponential part on the right-hand side: 

\begin{equation}
J_{\rm \gamma^{3} P2}=\frac{\Gp}{(\Ga\Gb)^{2}}\frac{1}{(\Gv^*\tau_{\rm lim})^{2}-1}\left[\left(1-\frac{\Ga^{2}+\Gb^{2}}{\Gp}\right)\frac{3}{\Gv^*}+\frac{\Gv^*}{\Gp}\right]\Hv(0)e^{-\Gv^*\tau}
\end{equation}
Applying the boundary condition defined by eq.~(\ref{eq::BC1JGb}) to the full solution $\JGb=J_{\rm\gamma^{3}P1}+J_{\rm\gamma^{3}P2}+J_{\rm\gamma^{3}H}$, we find $\Cd=0$.
The full solution of equation \eqref{eq::JGb} is hence given by:
\begin{equation}
\JGb(\tau)=J_{\rm \gamma^{3} P1}+J_{\rm \gamma^{3} P2}+\Cc e^{-\tau/\tau_{\rm lim}}
\label{eq::SolJGb}
\end{equation}
We can get an expression for the source function by replacing $\JG$ and $\JGb$ in the radiative equilibrium equation [eq.~(\ref{eq::radeqJG})]:
\begin{equation}
\begin{split}
B&=\Ca+3H\tau-\frac{(\Ga\Gb)^{2}}{\Gp}\Cc e^{-\tau/\tau_{\rm lim}}\\
   &+\frac{\left[3-(\Gv^*/\Ga)^{2}\right]\left[3-(\Gv^*/\Gb)^{2}\right]}{3\Gv^*(1-\Gv^{*2}\taulim^{2})}\Hv(0)e^{-\Gv^*\tau}
   \label{eq::BwithCste}
\end{split}
\end{equation}

 In order to get the complete solution of the problem, we need to determine the two remaining integration constants $\Ca$ and $\Cc$ using the boundary condition~(\ref{eq::BC2}). For that we need to calculate $\Ja(0)$, $\Jb(0)$, $\Ha(0)$ and $\Hb(0)$. The first two quantities can be evaluated using the values of $\JGa(0)$ and $\JGb(0)$ from equation \eqref{eq::SolJGa} and \eqref{eq::SolJGb} into the system \eqref{eq::JG2J}:
\begin{equation}
\begin{split}
\Ja(0)&=-\frac{\Ga\left(\Gb-1\right)}{\Ga-\Gb}\Ca\\ 
&+\frac{\Ga\left(3\left(\Ga+\Gb\right)\left(\Gb-1\right)-\Gv^{*2}+3\Ga^{2}\Gv^{*2}\taulim^{2}\right)}{\left(\Ga^{2}-\Gb^{2}\right)\Gv^*\left(\Gv^{*2}\taulim^{2}-1\right)}\Hv(0)\\
&-\frac{\Ga^{3}\Gb^{2}}{\Ga^{2}-\Gb^{2}}\Cc
\end{split}
\end{equation}
\begin{equation}
\begin{split}
\Jb(0)&=\frac{\Gb\left(\Ga-1\right)}{\Ga-\Gb}\Ca\\
&-\frac{\Gb\left(3\left(\Ga+\Gb\right)\left(\Ga-1\right)-\Gv^{*2}+3\Gb^{2}\Gv^{*2}\taulim^{2}\right)}{\left(\Ga^{2}-\Gb^{2}\right)\Gv^*\left(\Gv^{*2}\taulim^{2}-1\right)}\Hv(0)\\
&+\frac{\Gb^{3}\Ga^{2}}{\Ga^{2}-\Gb^{2}}\Cc
\end{split}
\end{equation}
Noting that:
\begin{equation}
\int_{0}^{\infty}E_{2}(\tau)d\tau=\frac{1}{2}
\end{equation}
\begin{equation}
\int_{0}^{\infty}\tau E_{2}(\tau)d\tau=\frac{1}{3}
\end{equation}
\begin{equation}
\int_{0}^{\infty}e^{-\alpha\tau}E_{2}(\tau)\,\mathrm{d}\tau=-\frac{1}{2}+\frac{1}{\alpha}-\frac{\ln(1+\alpha)}{\alpha^{2}}
\end{equation}
we can evaluate $\Ha(0)$ and $\Hb(0)$ by inserting eq.~(\ref{eq::BwithCste}) into eq.~(\ref{eq::H12}). Then, eq.~(\ref{eq::BC2}) is a linear system of two equations with two unknowns. After some calculations we get the expressions for $\Ca$ and $\Cc$:
\begin{equation}
\Ca=(a_{0}+a_{1}b_{0})H+(a_{1}b_{0}b_{1}(1+b_{2}+b_{3})+a_{2}+a_{3})\Hv(0)
\label{eq::first}
\end{equation}
\begin{equation}
\Cc=b_{0}H+b_{0}b_{1}(1+b_{2}+b_{3})\Hv(0)
\end{equation}
where we have:
\begin{equation}
a_0={1\over \gamma_1}+{1\over \gamma_2}
\label{eq::a0}
\end{equation}
\begin{equation}
a_1=-{1\over 3\tau_{\rm lim}^2}\left[{\gamma_{\rm p}\over 1-\gamma_{\rm p}}{\gamma_1+\gamma_2-2\over \gamma_1+\gamma_2}+(\gamma_1+\gamma_2)\tau_{\rm lim}-(A_{\rm t,1}+A_{\rm t,2})\tau_{\rm lim}^2\right]
\label{eq::a1}
\end{equation}
\begin{equation}
\begin{split}
a_{2}&=\frac{\taulim^{2}}{\Gp\Gv^{*2}}\times \\
&\frac{\left(3\Ga^{2}-\Gv^{*2}\right)\left(3\Gb^{2}-\Gv^{*2}\right)\left(\Ga+\Gb\right)-3\Gv^*\left(6\Ga^{2}\Gb^{2}-\Gv^{*2}\left(\Ga^{2}+\Gb^{2}\right)\right)}{1-\Gv^{*2}\taulim^{2}}
\end{split}
\end{equation}
\begin{equation}
a_{3}=-\frac{\taulim^{2}(3\Ga^{2}-\Gv^{*2})(3\Gb^{2}-\Gv^{*2})(A_{\rm{v,2}}+A_{\rm v,1})}{\Gp\Gv^{*3}(1-\Gv^{*2}\taulim^{2})}
\end{equation}
\begin{equation}
b_{0}=\left(\frac{\Ga\Gb}{\Ga-\Gb}\frac{A_{\rm t,1}-A_{\rm t,2}}{3}-\frac{(\Ga\Gb)^{2}}{\sqrt{3\Gp}}-\frac{(\Ga\Gb)^{3}}{(1-\Ga)(1-\Gb)(\Ga+\Gb)}\right)^{-1}
\end{equation}
\begin{equation}
b_{1}=\frac{\Ga\Gb(3\Ga^{2}-\Gv^{*2})(3\Gb^{2}-\Gv^{*2})\taulim^{2}}{\Gp\Gv^{*2}(\Gv^{*2}\taulim^{2}-1)}
\end{equation}
\begin{equation}
b_{2}=\frac{3(\Ga+\Gb)\Gv^{*3}}{(3\Ga^{2}-\Gv^{*2})(3\Gb^{2}-\Gv^{*2})}
\end{equation}
\begin{equation}
b_{3}=\frac{A_{\rm v,2}-A_{\rm v,1}}{\Gv^* (\Ga-\Gb)}
\end{equation}
where we defined:
\begin{equation}
A_{\mathrm{t},i}=\gamma_{i}^{2}\ln\left(1+\frac{1}{\taulim\gamma_{i}}\right)
\end{equation}
\begin{equation}
A_{\mathrm{v},i}=\gamma_{i}^{2}\ln\left(1+\frac{\Gv^*}{\gamma_{i}}\right)
\label{eq::Av}
\end{equation}

\subsection{Atmospheric temperature profile}
\label{sec::TemperatureProfile}

Using the relations $B=\sigma T^{4}/\pi$, $H=\sigma\tint^{4}/4\pi$ and $\Hv(0)=-\mu_{*}\sigma\tirr^{4}/4\pi$ and eq.~(\ref{eq::BwithCste}), we can derive the equation for the temperature at any optical depth:
\begin{equation}
\fbox{$
\begin{split}
T^4&={3T_{\rm int}^4\over 4} \left(\tau + A +B e^{-\tau/\tau_{\rm lim}}\right)\\
& +{3T_{\rm irr}^4\over 4}\mu_*\left(C+De^{-\tau/\tau_{\rm lim}}+Ee^{-\Gv^{*}\tau}\right)
\end{split}
$}
\label{eq::Tprofile}
\end{equation}
with
\begin{align}
\label{eq::A}
&A= {1\over3}(a_0+a_1b_0)\\
\label{eq::B}
&B= -{1\over3}{(\gamma_1\gamma_2)^2\over \gamma_{\rm p}}b_0\\
\label{eq::C}
&C=-{1\over3}\left[b_0 b_1 (1+b_2+b_3)a_1+a_2+a_3\right]\\
&D={1\over3}{(\gamma_1\gamma_2)^2\over \gamma_{\rm p}}b_0 b_1(1+b_2+b_3)\\
&E=\frac{\left[3-(\Gv^*/\Ga)^{2}\right]\left[3-(\Gv^*/\Gb)^{2}\right]}{9\Gv^*\left[(\Gv^*\taulim)^{2}-1\right]}
\label{eq::E}
\end{align}

\subsection{Grey limit}
In the grey limit, $\Gp\to1$ (as $\Ga$ and $\Gb$) and we obtain:
\begin{align}
 &A\to {2/3}\\
&B\to 0\\
&C\to{2}/{3}-{2}/{\Gv^{*2}}+{2}/{\Gv^*}+2\log(1+\Gv^*)\left({1}/{\Gv^{*3}}-{1}/{(3\Gv^*)}\right)\\
&D\to0\\
&E\to{\Gv^*}/{3}-{1}/{\Gv^*}
\end{align}
If we further assume that $\Gv^*\to 0$ we obtain $C\to{2}/{3}+{1}/{\Gv^*}$, $E\to -1/\Gv^*$ and the solution converges towards that of  \cite{Guillot2010} (see eq.~(\ref{eq::T4Guillot})). The fact that for other values of $\Gv^*$ our model differs from the solutions of \cite{Guillot2010} (see also \cite{Hansen2008}) is due to the different boundary conditions used in the two models as explained in Section~\ref{sec::BoundaryCondition}. However, calculations shows that the value of $C$ obtained here differ from the same coefficient extracted from eq.~(\ref{eq::T4Guillot}) by at most 12\% and that the two solutions converge also for $\Gv^{*}\to \infty$. As seen in figure \ref{fig::Guillot}, in the semi-grey limit, and when calculating the full temperature profile, our model differs by at most $2\%$ from the \cite{Guillot2010} model. The difference between the various solutions must be attributed to the Eddington approximation.

\subsection{Using the model}

The temperature vs. optical depth profile for our irradiated picket-fence model is given by eq.~(\ref{eq::Tprofile}). The profile has been derived using the Rosseland optical depth as vertical coordinate. It is therefore valid for any functional form of the Rosseland opacities. Equation \eqref{eq::tauP} allows to switch from $\tau$ to $P$ as the vertical coordinate. Although, for convenience, this expression contains 4 different variables, $\Ga,\Gb,\Gp,\taulim$, it must be kept in mind that, besides the Rosseland mean opacity, there are only two independent variables in the problem.
The variables $\beta$ and $\Rt\equiv\Ga/\Gb=\Ka/\Kb$ are the ones to consider to have a control on the opacity function used. The variables $\Gp$ and $\taulim$ are the ones to consider to have a control on the profile itself. $\Gp$ is directly related to the skin temperature of the planet (see section \ref{sec::Tskin}) whereas $\taulim$ is the optical depth at which the irradiated picket-fence model differs from the semi-grey model. The steps to use our model are as follows:
\begin{enumerate}[1)]
\item Choose the pair of variables suitable for the problem: ($\Rt$, $\beta$) or ($\Gp$, $\taulim$) for example
\item Using eqs~(\ref{eq::Gp}) to (\ref{eq::Gb}), calculate the values of $\Gp$, $\Ga$, $\Gb$ and $\taulim$
\item Using eqs~(\ref{eq::A}) to (\ref{eq::E}) and eqs~(\ref{eq::a0}) to (\ref{eq::Av}), calculate the coefficients A, B, C, D and E.
\item Using eq.~(\ref{eq::Tprofile}), calculate the temperature/optical depth profile
\item Using eq.~(\ref{eq::tauP}), calculate the pressure/optical depth relationship and therefore the pressure/temperature profile
\end{enumerate}
For Rosseland opacities depending on the temperature, step 5) can be iterated until convergence.
Given the apparent complexity of the solution, we provide a ready-to-use code\footnote[1]{https://www.oca.eu/parmentier/nongrey} in different languages that gives the temperature/optical depth profile (steps 1 to 4) or the temperature/pressure profile given a Rosseland mean opacity.

The relationship between the different variables are listed below:
\begin{equation}
\Gp=\beta+\Rt-\beta\Rt+\frac{\beta+\Rt-\beta\Rt}{\Rt}-\frac{\left(\beta+\Rt-\beta\Rt\right)^{2}}{\Rt}
\label{eq::Gp}
\end{equation}

\begin{equation}
\taulim=\frac{\sqrt{\Rt}\sqrt{\beta\left(\Rt-1\right)^{2}-\beta^{2}\left(\Rt-1\right)^{2}+\Rt}}{\sqrt{3}\left(\beta+\Rt-\beta\Rt\right)^{2}}
\end{equation}

\begin{equation}
\Rt=\frac{\sqrt{3\Gp}+3\Gp\taulim+\sqrt{\Delta}}{\sqrt{3\Gp}+3\Gp\taulim-\sqrt{\Delta}}
\end{equation}

\begin{equation}
\beta=\frac{\sqrt{\Delta}-\sqrt{3\Gp}+3\Gp\taulim}{2\sqrt{\Delta}}
\end{equation}

\begin{equation}
\Delta=3\Gp+3\sqrt{\Gp}\taulim\left(2\sqrt{3}\Gp+3\Gp^{3/2}\taulim-4\sqrt{3}\right)
\end{equation}

\begin{equation}
\Ga=\beta+\Rt-\beta\Rt
\end{equation}

\begin{equation}
\Gb=\frac{\beta+\Rt-\beta\Rt}{\Rt}
\end{equation}

\begin{equation}
\Ga=\frac{\sqrt{3\Gp}+3\Gp\taulim+\sqrt{\Delta}}{6\taulim}
\end{equation}
\begin{equation}
\Gb=\frac{\sqrt{3\Gp}+3\Gp\taulim-\sqrt{\Delta}}{6\taulim}
\label{eq::Gb}
\end{equation}

\begin{equation}
R=1+\frac{\Gp-1}{2\beta(1-\beta)}+\sqrt{\left(\frac{\Gp-1}{2\beta(1-\beta)}\right)^{2}+\frac{\Gp-1}{2\beta(1-\beta)}}
\end{equation}

\subsection{About averaging}

Equation~(\ref{eq::Tprofile}) can thus be considered as depending on $\Kr$, $\Gp\equiv \Kp/\Kr$ and $\beta$. While $\Kr$ can be considered as a function of pressure and temperature (e.g. extracted from a known Rosseland opacity table) when deriving the atmospheric temperature profile, it is important to realize that the analytical solution remains valid only if $\Gp$ and $\beta$ are held constant. This analytical solution therefore cannot accomodate consistent Rosseland {\em and\/} Planck opacities as a function of depth (a solution consisting of atmospheric slices with different values of $\Gp$ is derived in ~\citet{Chandrasekhar1935} for the non-irradiated case but becomes too complex to be handled easily). 

Furthermore, the solution is provided only for one fixed direction of the incoming irradiation. When considering the case of a non-resolved planet around a star, any information acquired on its atmosphere will have been averaged over at least a fraction of its surface. Solving this problem for the particular case of eq.~(\ref{eq::Tprofile}) goes beyond the scope of the present work, but it can be approximated relatively well on the basis of the study by \citet{Guillot2010}. This work shows that given an irradiation flux at the substellar point $\sigma T_{\rm sub}^4\equiv \sigma (R_*/D)^2T_*^4$, where $T_*$ is the star's effective temperature, $R_*$ its radius and $D$ the star-planet distance, the {\em average\/} temperature profile of the planet will be very close to that obtained from the 1-dimensional solution with an average angle $\mu^*=1/\sqrt{3}$ and an average irradiation effective temperature $T_{\rm irr}=(1-A)^{1/4} f^{1/4} T_{\rm sub}$, where $A$ is the (assumed) Bond albedo of the atmosphere and $f$ is a correction factor, equal to $1/4$ when averaging on the entire surface of the planet and equal to $1/2$ when averaging on the day-side only. This corresponds to the so-called ``isotropic approximation'' and is found to be within $2\%$ of the "exact" semi-grey average for a typical hot-Jupiter (see Fig.~2 of \citet{Guillot2010}). 

For the interpretation of spectroscopic and photometric data of secondary eclipses, the dayside average is often used ($f=1/2$). For the calculation of evolution models, the global average is the correct physical quantity to be used in the absence of a clear knowledge of the composition and opacity variations in latitude and longitude \citep[see][]{Guillot2010}. In that case, $f=1/4$ which is equivalent to setting the irradiation temperature equal to the usual equilibrium temperature defined as $T_{\rm eq}\equiv T_*(R_*/2D)^{1/2}$ \citep{Saumon1996}.  

Obviously however, detailed interpretations must use an approach mixing three-dimensional dynamical and radiative transfer models \citep[see][]{Guillot2010,Heng2012}. 
\subsection{Adding several bands in the visible}
\label{sec::MultipleBands}
Although, for the simplicity of the derivation, our model used only one spectral band in the visible channel, it can be easily extended to $n$ visible bands. The key point is that our equations, and in particular eq.~\eqref{eq::JGa} are linear in the visible. Thus, the equations can be solved for any linear combination of visible bands. In that case the first momentum of the visible intensity (see eq.~\eqref{eq::Jv}) would write :
\begin{equation}
J_{\rm v}(\tau)=-\frac{H_{\rm v}(0)}{\mu_{*}}\sum_{i=1}^{n}\beta_{\rm v \it i}e^{-\gamma_{\rm v\it i}^{*}\tau}
\end{equation}
Where $\beta_{\rm v\it i}$ is the relative spectral extend of the $i^{th}$ band and $\gamma_{\rm v\it i}=\kappa_{\rm v \it i}/\Kr$ with $\kappa_{\rm v\it i}$ the opacity in the $i^{th}$ visible band. Equation~\eqref{eq::Tprofile} then becomes: 
\begin{equation}
\fbox{$
\begin{split}
T^4&={3T_{\rm int}^4\over 4} \left(\tau + A +B e^{-\tau/\tau_{\rm lim}}\right)\\
& +\sum_{i=1}^{n}{3\beta_{\rm v\it i}T_{\rm irr}^4\over 4}\mu_*\left(C_{i}+D_{i}e^{-\tau/\tau_{\rm lim}}+E_{i}e^{-\gamma_{\rm v\it i}^{*}\tau}\right)
\end{split}
$}
\label{eq::TprofileMulti}
\end{equation}
where $C_{i}$, $D_{i}$ and $E_{i}$ are the coefficients $C$, $D$ and $E$ given by equation~\eqref{eq::C} to~\eqref{eq::E} where $\Gv^{*}$ have been replaced by $\gamma_{\rm v\it i}^{*}$.

\section{Comparisons}
\subsection{Comparison of non-irradiated solutions}

\begin{figure}[t!]
\centering
\includegraphics[width=\linewidth]{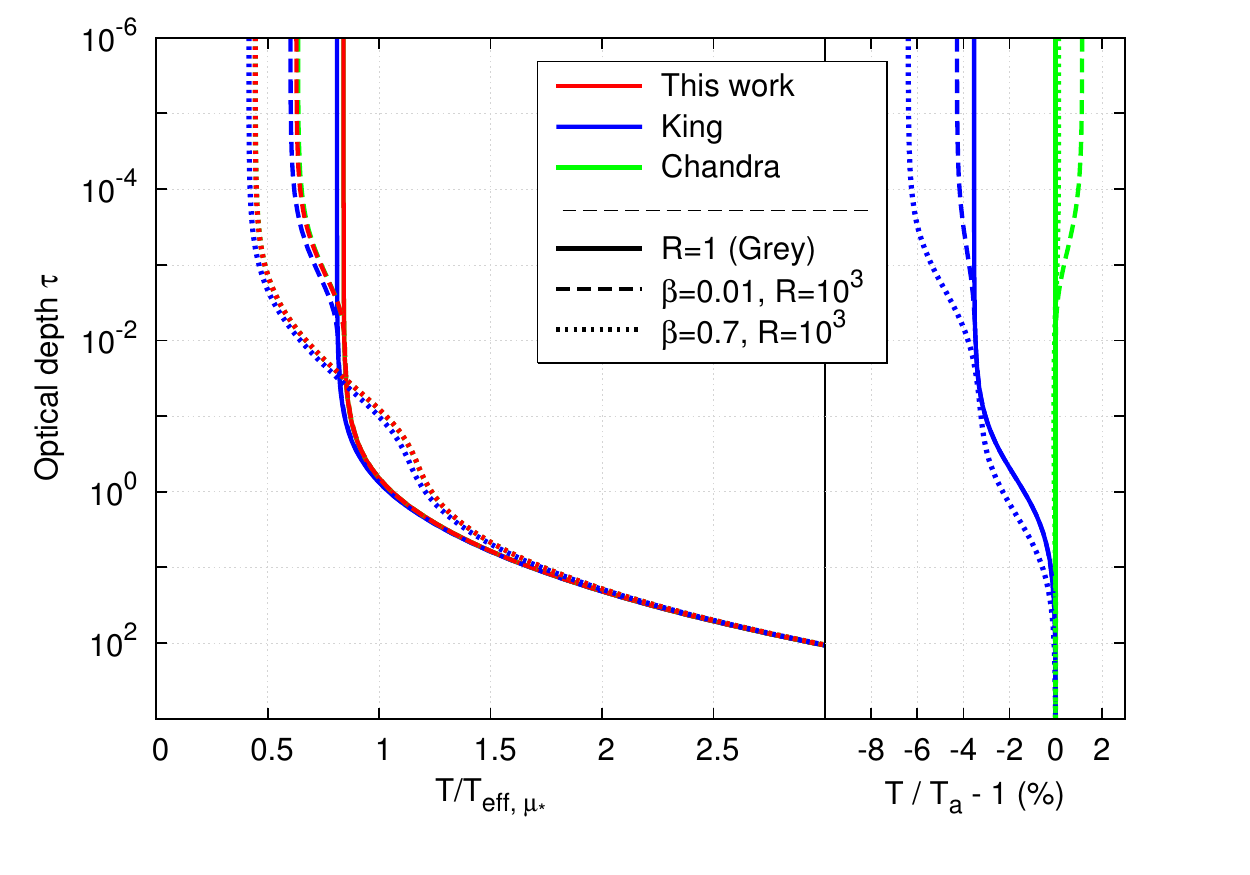}
\caption{Comparison of the non-irradiated solutions of the radiative transfer problem within the so-called picket-fence model approximation (see text). The left panel shows temperature (in $\teff$ units) versus optical depth. The right panel shows the relative temperature difference between our model and other works. The models shown correspond to the solutions of \citet{King1955} (blue lines), \citet{Chandrasekhar1935} (green lines), and this work (red). Different models correspond to the grey case (plain), i.e. $R=1$, and 2 non-grey cases: $\beta=0.01$, $R=10^{3}$ (dashed) and $\beta=0.7$, $R=10^{3}$ (dotted). ($R\equiv\kappa_1/\kappa_2$). The red and green lines are so similar that they are almost indistinguishable on the left panel.}
\label{fig::Chand}
\end{figure}

Figure~\ref{fig::Chand} shows a comparison between our results and the solutions of \citet{King1955} and \citet{Chandrasekhar1935}. The solutions are extremely close, the temperatures being always less than a few percent of each other. Our solution is almost identical to that of \citet{Chandrasekhar1935}, a consequence of using the Eddington approximation and similar boundary conditions. The difference of these with the exact solution from \citet{King1955} can be attributed to the Eddington approximation. 

The non-grey effects lead to colder temperatures at small optical depths. When $\beta$ is close to unity, a blanketing effect leads to a heating of the deeper layers too. All solutions have the correct behavior. 

\begin{figure}[t!]
\includegraphics[width=\linewidth]{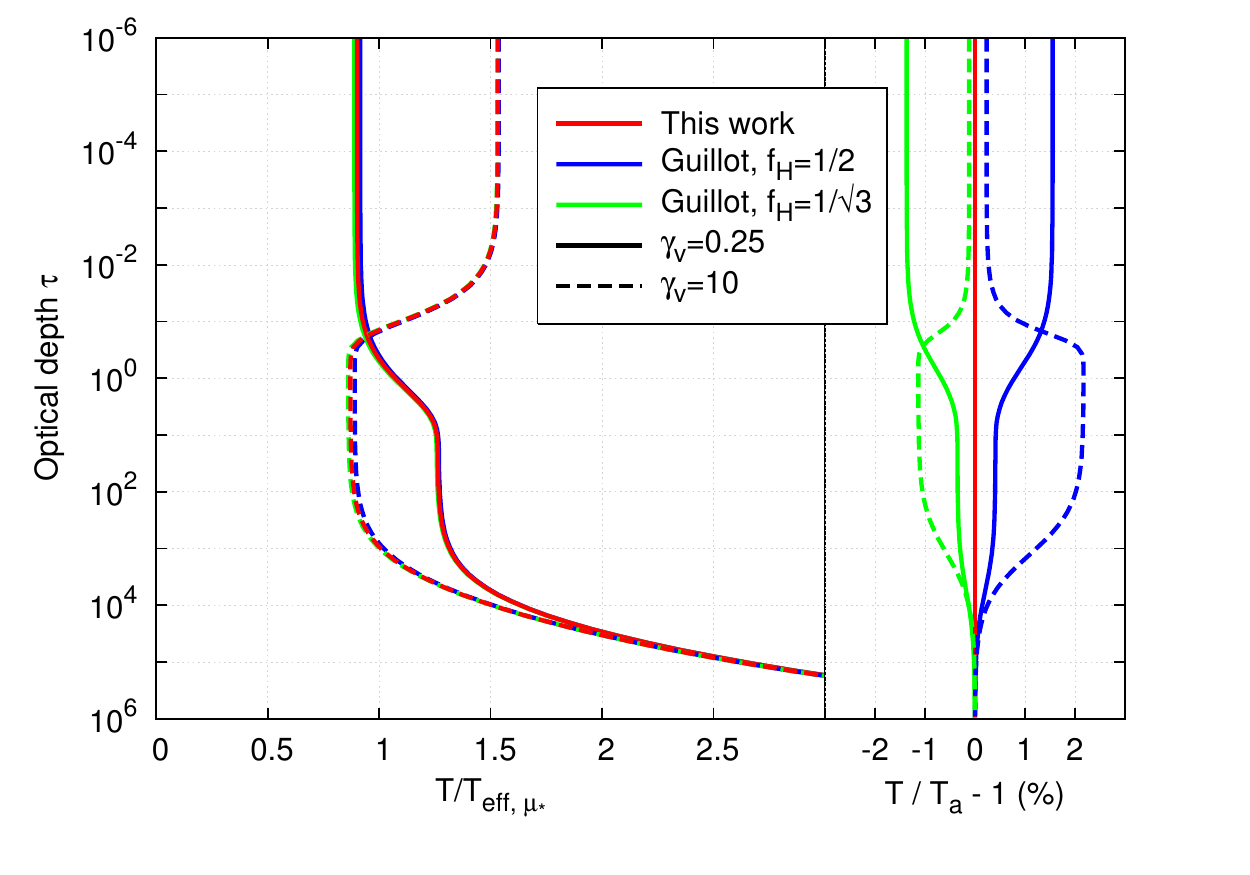}
\caption{Comparison between our model in the semi-grey limit and \citet{Guillot2010}. We used $\Gv=0.25$ (plain line) and $\Gv=10$ (dashed line). For the \citet{Guillot2010} model we show the curves for two different boundary conditions:$f_{\rm H}=1/2$ (blue) and  $f_{\rm H}=1/\sqrt{3}$ (green). We used $\mu_{*}=1/\sqrt{3}$.}
\label{fig::Guillot}
\end{figure}

\subsection{Comparison of irradiated solutions}

The solutions presented in this work for the irradiated case in the semi-grey case (i.e. $R\equiv \kappa_1/\kappa_2=1$) are very similar to those of \citet{Guillot2010}. As seen in Fig.~\ref{fig::Guillot}, the solutions obtained either with $f_H=1/2$, $f_H=1/\sqrt{3}$ have relative differences of up to $2\%$ with those of this work. These differences are of the same kind as those arising from the use of the Eddington approximation compared to exact solutions discussed previously. They are inherent to the approximation made on the angle dependence of the radiation field and implicitly linked to the choice of the different boundary solutions discussed in Section~\ref{sec::BoundaryCondition}.

\subsection{Comparison of skin temperatures}
\label{sec::Tskin}

\begin{figure}[h!]
\includegraphics[width=\linewidth]{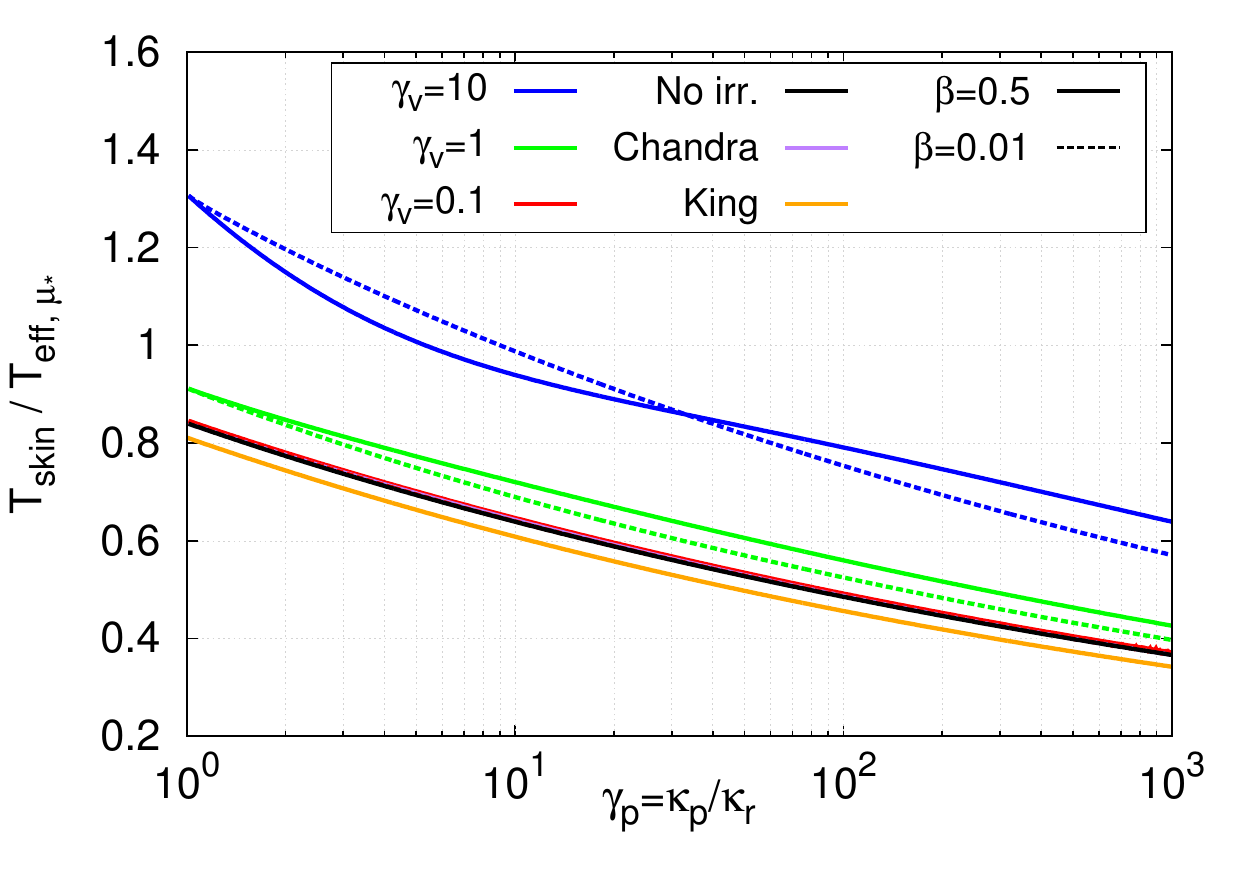}
\caption{Skin temperature of the planet given by our irradiated picket-fence model for different value of $\Gv$ and in the non-irradiated case. Curves for $\beta=0.01$ (plain lines) and $\beta=0.5$ (dash lines) are shown. Skin temperature from \citet{Chandrasekhar1935} and \citet{King1956} are also shown. For the irradiated case we used $\mu_{*}=1/\sqrt{3}$ and $f=0.5$. The $\Gv=0.1$, the non-irradiated and  the \citet{Chandrasekhar1935} curves are closely packed.}
\label{fig::Tskin}
\end{figure}

As discussed previously, the skin temperature (temperature at the limit of zero optical depth) is an important outcome of radiative transfer and in the case of non-irradiated models, an exact solution is available. We compare our results to analytical results in Fig.~\ref{fig::Tskin}. In the limit of a non-irradiated planet and in the limit $\Gv^{*}\to0$, our skin temperature converges to the one derived by \citet{Chandrasekhar1935}. This is an important test for the model, as for small values of $\Gv$, most of the stellar flux is absorbed in the deep layers of the planet and the model is expected to behave as a non-irradiated model with the same effective temperature. Moreover, we note that for small values of $\Gv$, the skin temperature is affected only by $\Gp$ as was already claimed by \citet{King1956} and \citet{Chandrasekhar1935}. This conclusion no longer applies for larger values of $\Gv$ for which the skin temperatures then also depend on $\beta$. This can be seen by comparing the dotted lines and plain lines of the same color in figure \ref{fig::Tskin}. At a given value of $\Gp$, a larger value of $\beta$ corresponds to a smaller $\kappa_{\rm 2}/\kappa_{\rm 1}$. Depending of the value of $\beta$, the stellar irradiation can be absorbed in a region which can be optically thick to the two thermal bands, only one, or none, leading to a different behavior for the skin temperature.

\section{Consequences of non-grey effects}

In this section we study the physical processes that shape our non-grey temperature profile. To overcome the apparent complexity of our solution, we first derive an approximate expression for the thermal fluxes at the top of the atmosphere. Using this expression, we obtain a much simpler expression for the skin and the deep temperatures. Comparing these expressions with their semi-grey equivalent, we get physical insights on the processes that shape the temperature profile.

\subsection{Estimation of the fluxes in the different bands}

In steady state, all the energy that penetrates the atmosphere must be radiated away. Thus, the radiative equilibrium at the top of the atmosphere provides great insights on the physical processes that shape the temperature profile. In particular whether the thermal fluxes are transported by the channel of highest opacity (channel 1) or the channel of lowest opacity (channel 2) is of particular importance.

As seen in eq.~\eqref{eq::Tprofile} the contribution to the final temperature of the internal luminosity and of the external irradiation are independent. Thus, the thermal fluxes can be split in two independent contributions that can be studied separatly:
\begin{equation}
H_{i}=H_{i \rm,\,int}+H_{i \rm,\,irr}
\end{equation}
Figure~\ref{fig::FluxGv} shows which thermal band actually carries the thermal flux $H_{\rm irr}$ out of the atmosphere. This depends strongly whether the stellar irradiation is absorbed in the upper or in the deep atmosphere. If it is deposited in the deep layers of the planet (\ie $\Gv<<1$), most of the flux is transported by the second thermal channel {\it whatever the width of the second channel}. Conversely, when the stellar irradiation is deposited in the upper atmosphere, most of the flux is carried by the first thermal channel {\it whatever the width of the first channel}. The tipping point, \ie~when each channel carries half of the flux, is reached when $\Gv=\taulim^{-1}$. Fig.~\ref{fig::TauLim} shows the variations of $\taulim$ with the width and the strength of the two thermal opacity bands. $\taulim$ increases with $\beta$ but decreases with $\Ka/\Kb$. It always corresponds to an optical depth where the first channel is optically thick and the second is optically thin.

For large values of $\Gp$ (\ie~$\Gp>2$), we can approximate the ratio of the thermal fluxes related to the irradiation by a much simpler expression:
\begin{equation}
\label{eq::H2oH1Approx}
\frac{H_{\rm 1,\, irr}(0)}{H_{\rm 2,\, irr}(0)}\approx\frac{\beta}{\sqrt{\Gp}}+\frac{1}{\frac{1-\beta}{\sqrt{\Gp}}+\frac{1}{\Gv^{*}\taulim}}
\end{equation}
As shown in Fig.~\ref{fig::FluxGv} this expression matches correctly the expression for the analytical model. Depending on the value of $\Gv^{*}\taulim$ the expression reduces to:
\begin{subequations}\label{eq::Cond}
\label{eq::Cond12}
\begin{equation}
   \frac{H_{\rm 1,\, irr}(0)}{H_{\rm 2,\, irr}(0)}\approx\frac{\beta}{\sqrt{\Gp}} + \Gv^{*}\taulim, \text{     when } \Gv^{*}\taulim< 1 \label{eq::Cond1}
     \end{equation}
    \begin{equation}
    \frac{H_{\rm 2,\, irr}(0)}{H_{\rm 1,\, irr}(0)}\approx  \frac{1-\beta}{\sqrt{\Gp}} + \frac{1}{\Gv^{*}\taulim},  \text{    when } \Gv^{*}\taulim> 1 \label{eq::Cond2}
    \end{equation}
\end{subequations}

We now look for a similar expression for the thermal fluxes resulting from the internal luminosity ($H_{\rm int}$). Because the internal luminosity irradiates the atmosphere from below, the resulting thermal fluxes behave similarly to the irradiated when $\Gv\to0$, thus we have:
\begin{equation}
\label{eq::H2oH1ApproxWithHint}
\frac{H_{\rm 1,\, int}(0)}{H_{\rm 2,\, int}(0)}\approx\frac{\beta}{\sqrt{\Gp}}
\end{equation}
As $\Gp$ is always greater than, the internal luminosity is always transported by channel 2, the channel of lowest opacity.

\subsection{The skin temperature}

The skin temperature reveals the behavior of the atmosphere at low optical depths. This is the part of the atmosphere probed during the transit of an exoplanet in front of its host star and is therefore of particular importance to interpret the observations. 
Figure~\ref{fig::DTskin} shows that in the irradiated case non-grey effects always tend to lower the skin temperature compared to the semi-grey case. This upper atmospheric cooling is already significant ($>10\%$) for slightly non-grey opacities (\ie~$\Gp\approx2$). For larger values of $\Gp$ the cooling is stronger, reaching $50\%$ for $\Gp\approx10-1000$.  Conversely to the non-irradiated case, the skin temperature is not only a function of $\Gp$ but also depends on $\beta$, \ie~not only the mean opacities are relevant but also their actual shape. For large values of $\beta$, when the stellar irradiation is absorbed in the upper layers of the atmosphere (\eg~$\Gv=100$) the cooling is more efficient than when the stellar irradiation is absorbed in the deep layers (\eg~$\Gv=0.01$) whereas for small values of $\beta$ the cooling is independent on $\Gv$.

The skin temperature results directly from the radiative equilibrium of the upper atmosphere. Using the boundary condition~\eqref{eq::BC2} in the radiative equlibrium equation~\eqref{eq::RadEq} evaluated at $\tau=0$ we can write:
\begin{equation}
2\Ga\Ha(0)+2\Gb\Hb(0)-\Gv^{*}\Hv(0)=\Gp B(0)
\label{eq::RadEq2}
\end{equation}
where the skin temperature is given by $\tskin^{4}=\pi B(0)/\sigma$. The skin temperature, depends on the values of $\Ha(0)$ and $\Hb(0)$ and thus on whether the stellar irradiation is absorbed in the deep atmosphere or in the upper atmosphere.

\subsubsection{Case of a deep absorption of the irradiation flux} 

When $\Gv^{*}\taulim<1$, the stellar irradiation is absorbed in the deep layers of the atmosphere, where the second thermal band, the band of lowest opacity, is optically thick. Thus, most of the flux is transported by the second thermal band and we have $\Hb(0)=H_{\infty}-\Hv(0)$. For large values of $\Gp$, using eq.~\eqref{eq::Cond1} and eq.~\eqref{eq::H2oH1ApproxWithHint} we get $\Ga\Ha(0)/\Gb\Hb(0)>\sqrt{\Gp}/(1-\beta)$ which is always larger than one. Thus, although most of the flux is in the second thermal band, it is the first band, the band of highest opacity, that sets the radiative equilibrium. Neglecting the second term in eq~\eqref{eq::RadEq2} and calculating $\Ha(0)$ with eqs.~\eqref{eq::Cond1} and~\eqref{eq::H2oH1ApproxWithHint} we obtain:
\begin{equation}
B(0)=\frac{2\Ga\beta}{\Gp\sqrt{\Gp}}H_{\infty}-\left(\frac{2\Ga\beta}{\Gp\sqrt{\Gp}}+\frac{2\Gv^{*}\taulim\Ga}{\Gp}+\frac{\Gv^{*}}{\Gp}\right)\Hv(0)
\end{equation}
Noting that for large values of $\Gp$, $\taulim\approx\beta(1-\beta)^{-1}\left(\sqrt{3\Gp}\right)^{-1}$ and $\Ga\approx\Gp/\beta$, the equation becomes:
\begin{equation}
B(0)=\frac{2}{\sqrt{\Gp}}H_{\infty}-\left(\frac{2}{\sqrt{\Gp}}+\frac{2\Gv^{*}}{(1-\beta)\sqrt{3\Gp}}+\frac{\Gv^{*}}{\Gp}\right)\Hv(0)
\end{equation}
Replacing the fluxes by their equivalent temperature we get an expression for the skin temperature valid for $\Gv^{*}\taulim<1$ and $\Gp>2$:
\begin{equation}
\tskin^{4}=\frac{2}{\sqrt{\Gp}}\frac{\tint^{4}+\mu_{*}\tirr^{4}}{4}+\left(\frac{2\Gv^{*}}{(1-\beta)\sqrt{3\Gp}}+\frac{\Gv^{*}}{\Gp}\right)\frac{\mu_{*}\tirr^{4}}{4}
\end{equation}
When $\Gv^{*}\taulim<1$, the first term dominates and the expression differs by a factor $1/\sqrt{\Gp}$ from the semi-grey case (eq.~\eqref{eq::Tskin-Guillot}). Because $\Gp>1$ for non-grey opacities, the skin temperature is always smaller in the non-grey case than in the grey case, as shown in Fig.~\ref{fig::DTskin}. \\\\
\emph{Physical interpretation.}
When $\Gv^{*}\taulim<1$ most of the irradiation is absorbed where both thermal channels are optically thick. The flux is mainly transported by the channel of lowest opacity $\Kb$ but only the residual flux transported by the channel of highest opacity $\Ka$ contributes to the radiative equilibrium at the top of the atmosphere. Because it represents only a small part of the total flux, the upper atmospheric temperatures are smaller than in the semi-grey case. The larger the departure from the semi-grey opacities, the cooler the skin temperature, without lower bound.

\subsubsection{Case of a shallow absorption of the irradiation flux} 

\begin{figure}[t!]
\includegraphics[width=\linewidth]{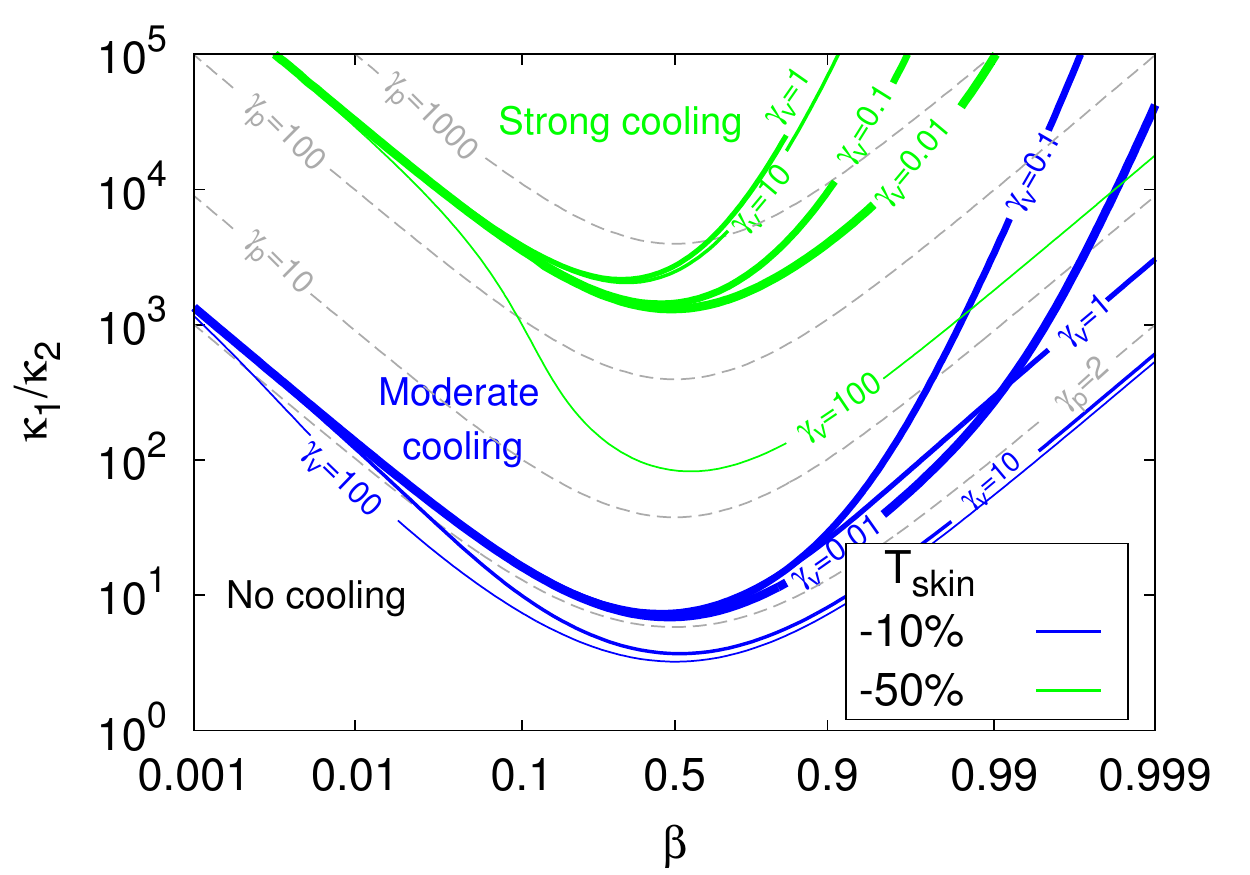}
\caption{Contours of the relative difference between the skin temperature in the non-grey model and in the semi-grey model for different values of $\Gv$ in function of the width of the lines and their strength. The non-grey atmosphere is $10\%$ (resp. $50\%$) cooler than the semi-grey atmosphere above the blue (resp. green) lines. The dashed lines are contours of $\Gp$. We used $\mu_{*}=1/\sqrt{3}$.}
\label{fig::DTskin}

\includegraphics[width=\linewidth]{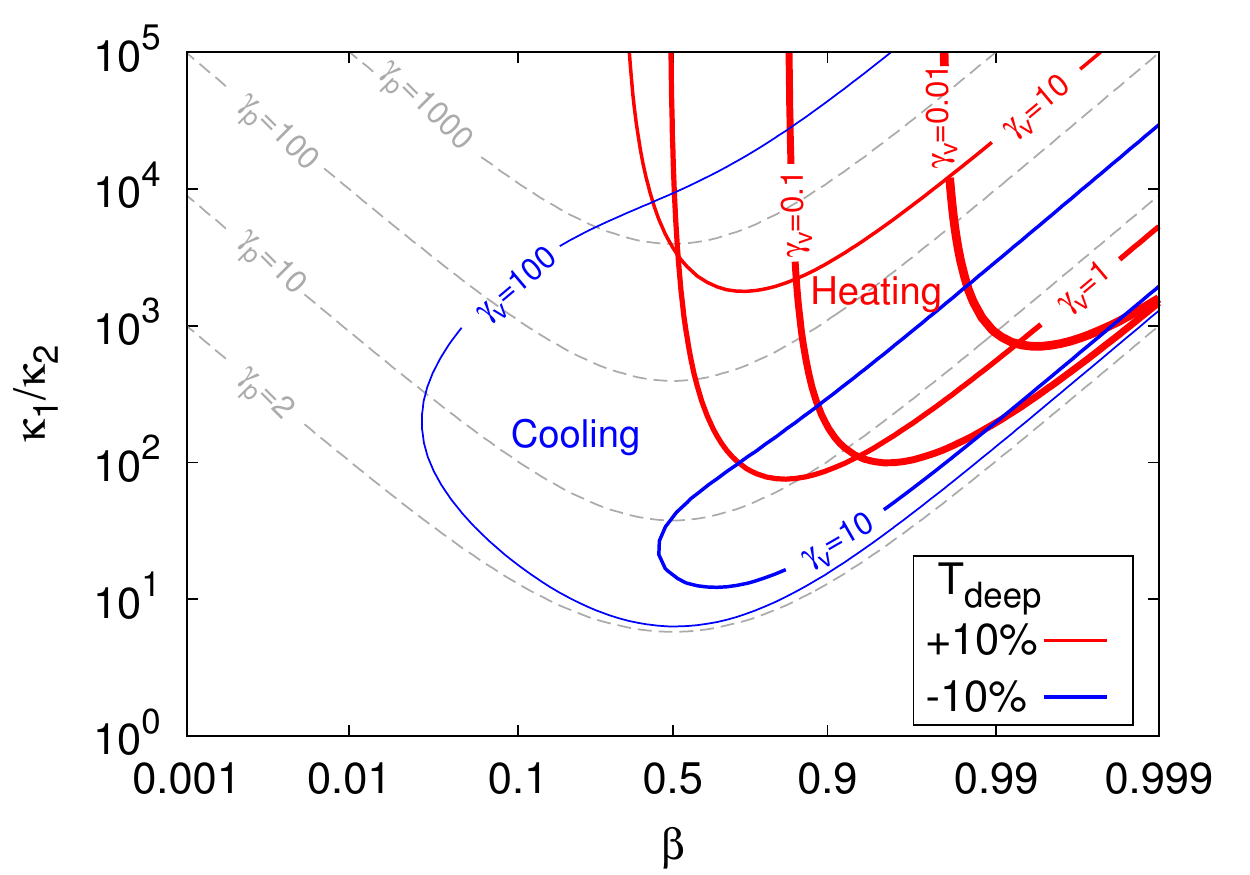}
\caption{Contours of the relative difference between the skin temperature in the non-grey model and in the semi-grey model for different values of $\Gv$ in function of the width of the lines and their strength. The non-grey atmosphere is $10\%$ hotter (resp. cooler) than the semi-grey atmosphere inside the red (resp. blue) contours. The dashed lines are Contours of $\Gp$. We used $\mu_{*}=1/\sqrt{3}$.}
\label{fig::DTdeep}
\end{figure}

When $\Gv^{*}\taulim>1$, most of the stellar irradiation is absorbed in the upper atmosphere, where only the first thermal band is optically thick. According to.~\eqref{eq::Cond1}, most of the flux originating from the irradiation $H_{\rm irr}$ is carried by the first thermal band, the band of highest opacity. Conversely, following eq.~\eqref{eq::H2oH1ApproxWithHint}, the internal luminosity is still transported by the second thermal channel, as in the $\Gv\taulim<1$ case. Thus, the radiative equilibrium of the upper atmosphere is still determined by the channel of highest opacity, channel 1 and the second term of eq.~\eqref{eq::RadEq2} can be neglected. Conversely to the case $\Gv\taulim<1$, the top boundary condition now reads $\Ha(0)\approx H_{\rm 1,\, int}-\Hv(0)$. Using eq.~\eqref{eq::H2oH1ApproxWithHint} to calculate $H_{\rm 1,\, int}$ and noting that for large values of $\Gp$, $\Gp\approx\beta\Ga$, the radiative equilibrium becomes:
\begin{equation}
B(0)=\frac{2}{\sqrt{\Gp}}H_{\infty}-\left(\frac{2}{\beta}+\frac{\Gv^{*}}{\Gp}\right)\Hv(0)
\end{equation}
Replacing the fluxes by their equivalent temperatures we get an expression for the skin temperature valid for $\Gv^{*}\taulim>1$ and $\Gp>2$:
\begin{equation}
\tskin^{4}=\frac{2}{\sqrt{\Gp}}\frac{\tint^{4}}{4}+\left(\frac{2}{\beta}+\frac{\Gv^{*}}{\Gp}\right)\frac{\mu_{*}\tirr^{4}}{4}
\end{equation}
This relation differs from the case $\Gv^{*}\taulim<1$ as the factor $1/\sqrt{\Gp}$ before the irradiation temperatures is replaced by a factor $1/\beta$. Thus, the skin temperature cannot become arbitrarily low anymore. However, for large values of $\Gv$, the second term in the parenthesis dominates and the skin temperature decreases proportionally to $1/\sqrt{\Gp}$, which is faster than in the case $\Gv^{*}\taulim<1$. As an example, in Fig~\ref{fig::DTskin}, for $\Gv=100$, the skin temperature decreases much faster when $\Gp$ increases for large values of $\beta$, \ie~when $\Gv\taulim>1$.\\\\
\emph{Physical interpretation.}
When $\Gv\taulim>1$, most of the incident irradiation is absorbed in the upper atmosphere, where the second channel is optically thin. Therefore it is mainly transported by the channel of highest opacity: channel 1. Similarly to the case $\Gv^{*}\taulim<1$, the radiative equilibrium at the top of the atmosphere is set by the channel of highest opacity, the one that carries most of the thermal flux. Therefore all the flux from the irradiation contributes to the radiative equilibrium of the upper layers and the skin temperature cannot cool as much as in the $\Gv^{*}\taulim<1$ case, its lowest value being $\mu^{*}\tirr^{4}/2\beta$. However, for large values of $\Gv$ and as long as long $\Gp<\Gv$, $\tskin^{4}$ decrease faster than in the case $\Gv\taulim>1$. This confines the stratosphere (\ie the atmospheric levels with a temperature inversion) around the $\tau=\taulim$ level whereas it extends up to $\tau=0$ in the semi-grey case (see Figs.~\ref{fig::AllProfiles1},~\ref{fig::AllProfiles2} and \ref{fig::AllProfiles3} hereafter).

\begin{figure}[t!]

\includegraphics[width=\linewidth]{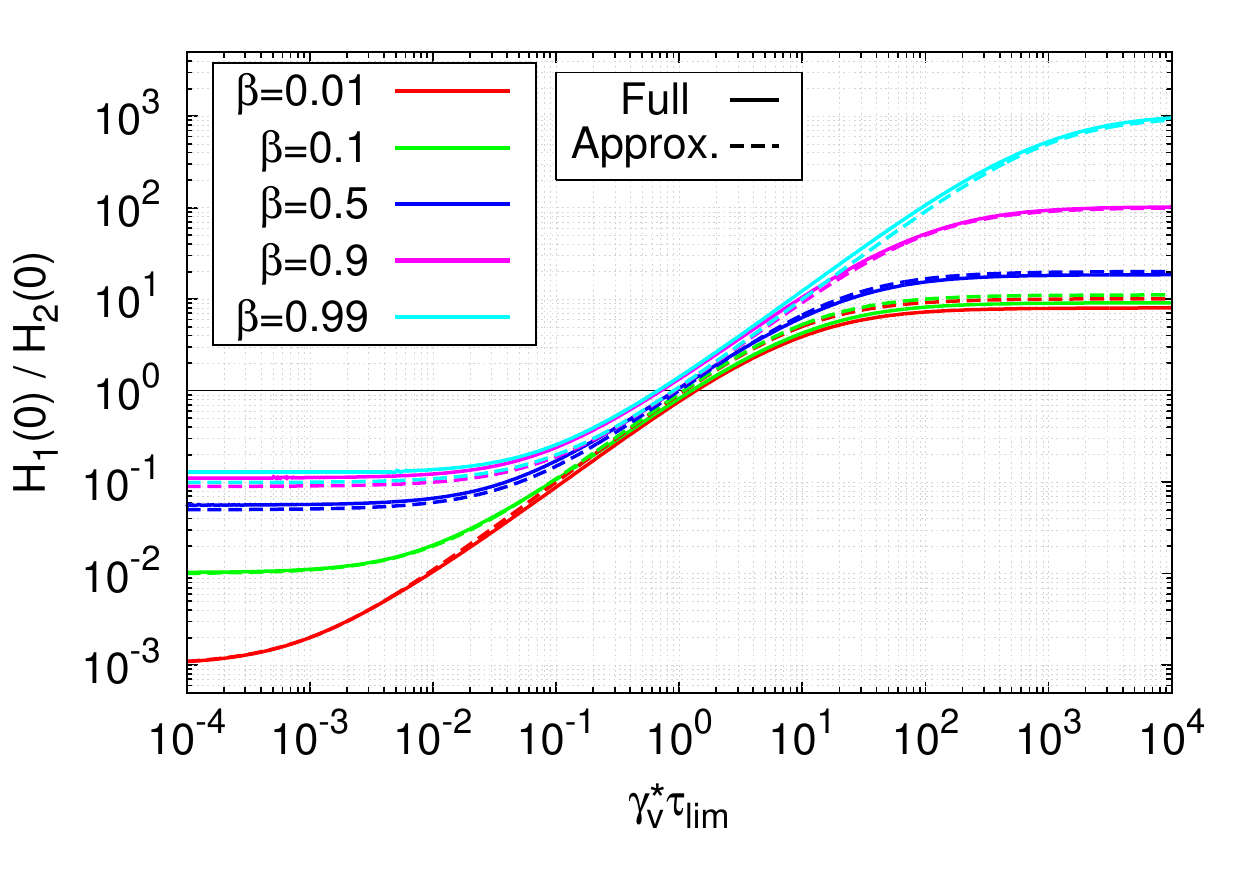}
\caption{Ratio of the total flux in the two thermal bands in function of $\Gv^{*}\taulim$ for different $\beta$ and for $\Gp=100$ given by our analytical model (plain) and by the approximate expression~\eqref{eq::H2oH1Approx} (dashed). We used $\mu_{*}=1/\sqrt{3}$ and $\tint=0$.}
\label{fig::FluxGv}

\includegraphics[trim=0cm 0cm 0cm 0cm, clip=true, width=\linewidth]{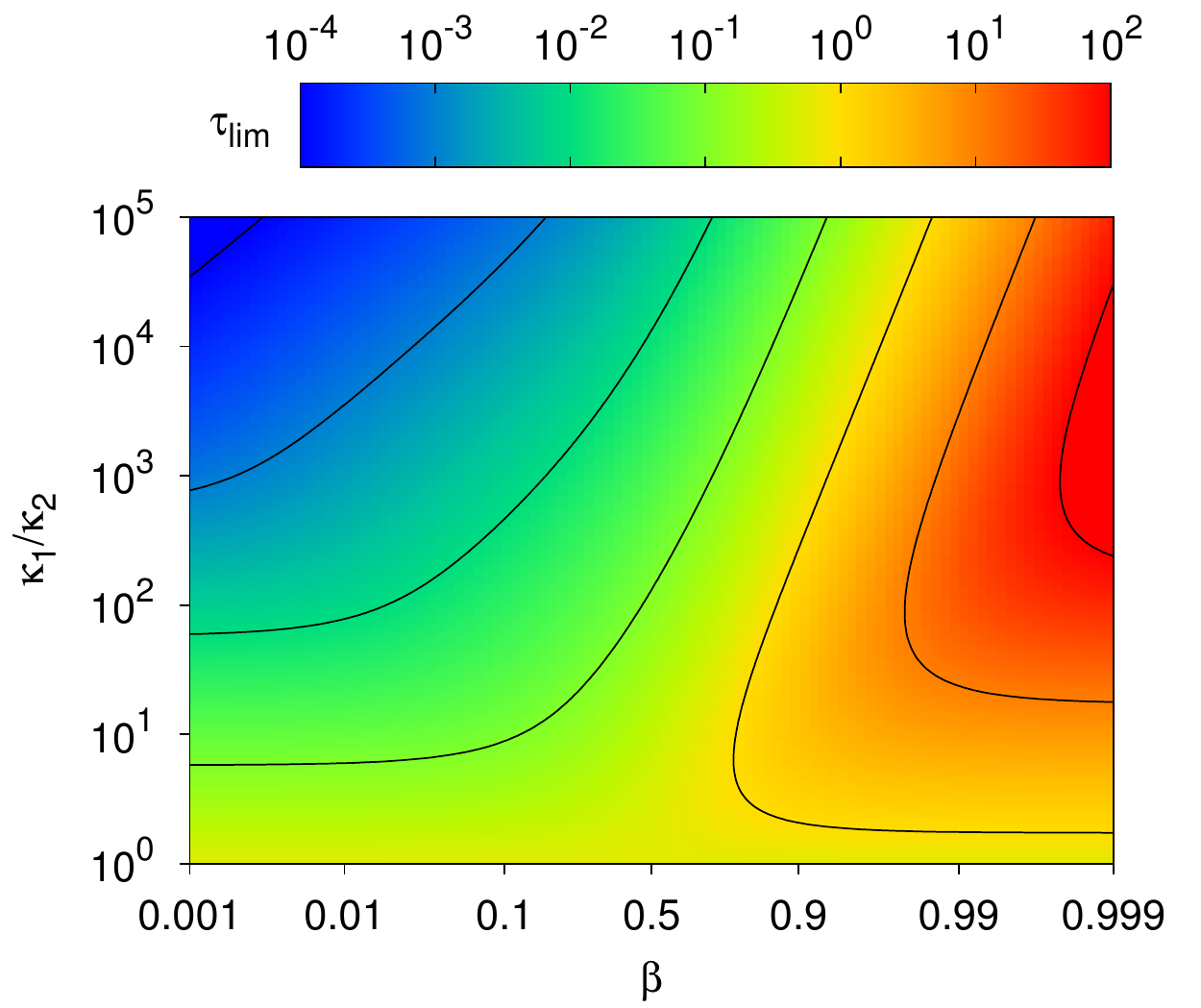}
\caption{Value of $\taulim$ in function of the width of the lines $\beta$ and their strength $\Ka/\Kb$. The x-axis is in logit scale, where the function logit is defined as $\rm{logit}(x)=\log(x/(1-x))$}
\label{fig::TauLim}
\end{figure}

\subsection{The deep temperature}

The temperature of the deep atmosphere is a fundamental outcome from radiative transfer models as it reveals the energy exchange between the planet and its surroundings. Therefore it is often used as a boundary condition of planetary interior models. We define the deep temperature as :
\begin{equation}
T_{\rm deep}^{4}=\lim_{\tau \to \infty}T(\tau)^{4}-3\tint^{4}\tau
\label{eq::Tdeep}
\end{equation}
Thus, the temperature of the deep atmosphere can be well approximated as $T(\tau)^{4}=\tdeep^{4}+3\tint^{4}\tau$ between the $\tau\approx1$ level and the radiative/convective boundary. For irradiated planets, the deep temperature corresponds to the isothermal zone around $\tau\approx1$. As seen in Figure~\ref{fig::DTdeep}, it as a complex behavior. For small values of $\Gv$ the deep temperature increases compared to the semi-grey case whenever $\beta$ becomes large enough; an effect known as the {\it line blanketing effect} in the stellar literature~\citep[see][for example]{Milne1921,Chandrasekhar1935,Hubeny1995}. This effect is always maximum when $\Gv^{*}\approx\taulim^{-1}$ (see hereafter Fig.~\ref{fig::TauLim}). Conversely, for large values of $\Gv^{*}$ (\ie~$\Gv^{*}>10$), the deep atmosphere warms up only for large values of $\Gp$ ($\Gp>\Gv^{2}$) whereas it becomes cooler than in the semi-grey case for smaller values of $\Gp$, a behavior that was not spotted in previous analytical models.

The deep atmospheric temperature is directly set by the boundary condition at the top of the atmosphere. From eq.~\eqref{eq::BwithCste}, we see that when $\tau\to\infty$ :
\begin{equation}
T_{\rm deep}^{4}=\lim_{\tau \to \infty}B(\tau)-3H_{\infty}\tau=\Ca
\end{equation}
where $\Ca$ is set by the top boundary condition~\eqref{eq::BC2} applied on $\JG(0)$ (see eq.~\eqref{eq::BCJG}):
\begin{equation}
2\frac{\Ha(0)}{\Ga}+2\frac{\Hb(0)}{\Gb}=\Ca+\frac{3}{\Gv^{*}}\Hv
\end{equation}
Similarly to the skin temperature, the deep temperature depends on $\Ha(0)$ and $\Hb(0)$ and depends whether the thermal flux is transported by the first or by the second thermal channel, \ie~whether $\Gv\taulim$ is larger or smaller than one.

\subsubsection{Case of a deep absorption of the irradiation flux} 

In the case $\Gv\taulim<1$ most of the thermal flux is transported by the second thermal channel and because $\Ga>>\Gb$ we can write:
\begin{equation}
2\frac{\Hb(0)}{\Gb}\approx\Ca+\frac{3}{\Gv^{*}}\Hv
\label{eq::WithH2}
\end{equation}
Applying the radiative equilibrium at the top of the atmosphere, and considering that most of the flux is carried by the second thermal channel, we get:
\begin{equation}
\Hb(0)\approx H_{\infty}-\Hv(0)
\end{equation}
Thus, we can calculate $\Ca$ and obtain :
\begin{equation}
B(\tau)-3H_{\infty}\tau\underset{\tau\to\infty}{\sim}\frac{2}{\Gb}H_{\infty}-\left(\frac{2}{\Gb}+\frac{3}{\Gv^{*}}\right)\Hv(0)
\end{equation}
For large values of $\Gp$ and thus large value of $\Gp$, $\Gb\approx(1-\beta)$. Replacing the fluxes by their equivalent temperatures we get an expression for $T_{\rm deep}$ valid for $\Gv^{*}\taulim<1$ and $\Gp>2$ :
\begin{equation}
T_{\rm deep}^{4}\approx\frac{2}{1-\beta}\frac{\tint^{4}+\mu_{*}\tirr^{4}}{4}+\frac{3}{\Gv^{*}}\frac{\mu_{*}\tirr^{4}}{4}
\label{eq::Tdeep1}
\end{equation}
This expression differs from the semi-grey value of~\citet{Guillot2010} by a factor $1/(1-\beta)$ multiplying the first term. Thus, when $\beta\to1$, the temperature becomes warmer than in the semi-grey case, as seen for the small values of $\Gv$ in Fig.~\ref{fig::DTdeep}.\\\\
\emph{Physical interpretation.} When $\Gv^{*}\taulim<1$, most of the flux from the star is absorbed in the deep atmosphere and is principally transported by the channel of lowest opacity (channel 2), even when the width of this channel is smaller than the width of the first thermal channel. Whenever $\beta\to1$, the width of the second channel decreases. In order to keep transporting most of the thermal flux, the flux per wavelength in the second channel must increase. This increases the temperature where the second channel is optically thick, \ie~in the deep atmosphere. This is equivalent to the \emph{line blanketing effect} well studied in stars \citep[see][for example]{Milne1921,Chandrasekhar1935,Hubeny1995}.

\subsubsection{Case of a shallow absorption of the irradiation flux} 

When $\Gv^{*}\taulim>1$, we have $H_{\rm 1,\,irr}(0)>>H_{\rm 2,\,irr}(0)$. Moreover, using  eq.~\eqref{eq::Cond2} we can show that $(H_{\rm 2,\,irr}(0)/\Gb)/(H_{\rm 1,\,irr}(0)/\Ga)>\sqrt{\Gp}/\beta$ which is larger than 1. Thus eq.~\eqref{eq::WithH2} remains valid. However, conversely to the case $\Gv^{*}\taulim<1$, the top boundary condition now reads :
\begin{equation}
\begin{aligned}
\Ha(0)&\approx -\Hv(0)+H_{\rm 2,\, int}\\
\Hb(0)&\approx H_{\infty}+H_{\rm 2,\, irr}
\end{aligned}
\end{equation}
Where $H_{\rm 2,\, irr}$ is given by eq.~\eqref{eq::Cond2} and $H_{\rm 1,\,int}$ by eq.~\eqref{eq::H2oH1ApproxWithHint}. This leads to:
\begin{equation}
C_{\rm1}=\frac{2}{\Gb}\frac{1-\beta}{\sqrt{\Gp}}(H_{\infty}-\Hv(0))-\frac{2}{\Gb\Gv^{*}\taulim}\Hv(0)-\frac{3}{\Gv}\Hv(0)
\end{equation}
Again, for large values of $\Gp$, $\Gb\to1-\beta$ and replacing the fluxes by their equivalent temperatures we get an expression for $T_{\rm deep}$ valid for $\Gv^{*}\taulim>1$ and $\Gp>2$
\begin{equation}
T_{\rm deep}^{4}=\frac{2}{1-\beta}\frac{\tint^{4}}{4}+\frac{2}{\sqrt{\Gp}}\frac{\mu^{*}\tirr^{4}}{4}+\left(\frac{3}{\Gv}+\frac{2}{1-\beta}\frac{1}{\Gv^{*}\taulim}\right)\frac{\mu^{*}\tirr^{4}}{4}
\end{equation}
When $\Gv^{*}\taulim>>1$, the contribution to the deep temperature of the irradiation temperature becomes inversely proportional to $\sqrt{\Gp}$. As $\Gp>1$, the deep temperature is smaller in the non-grey case than in the semi-grey case. This is illustrated by the cases $\Gv=10$ and $\Gv=100$ of Fig.~\ref{fig::DTdeep}. When $\Gv^{*}\taulim\to1$, the term in $1/\sqrt{\Gp}$ becomes very small compared to the term in $1/(1-\beta)$ and the expression converges toward equation eq.~\eqref{eq::Tdeep1}, valid for $\Gv^{*}\taulim<1$. \\\\
\emph{Physical interpretation.} 
When $\Gv^{*}\taulim>>1$, the incident irradiation is absorbed in the upper atmosphere, where only the channel of highest opacity is optically thick. Thus, the channel of highest opacity $\Ka$ transports all the energy and radiates it directly to space. The incident irradiation is not transported to the deep atmosphere, leading to a cooler deep atmosphere.

\subsection{Outgoing flux}

\begin{figure}[!t]
\includegraphics[width=\linewidth]{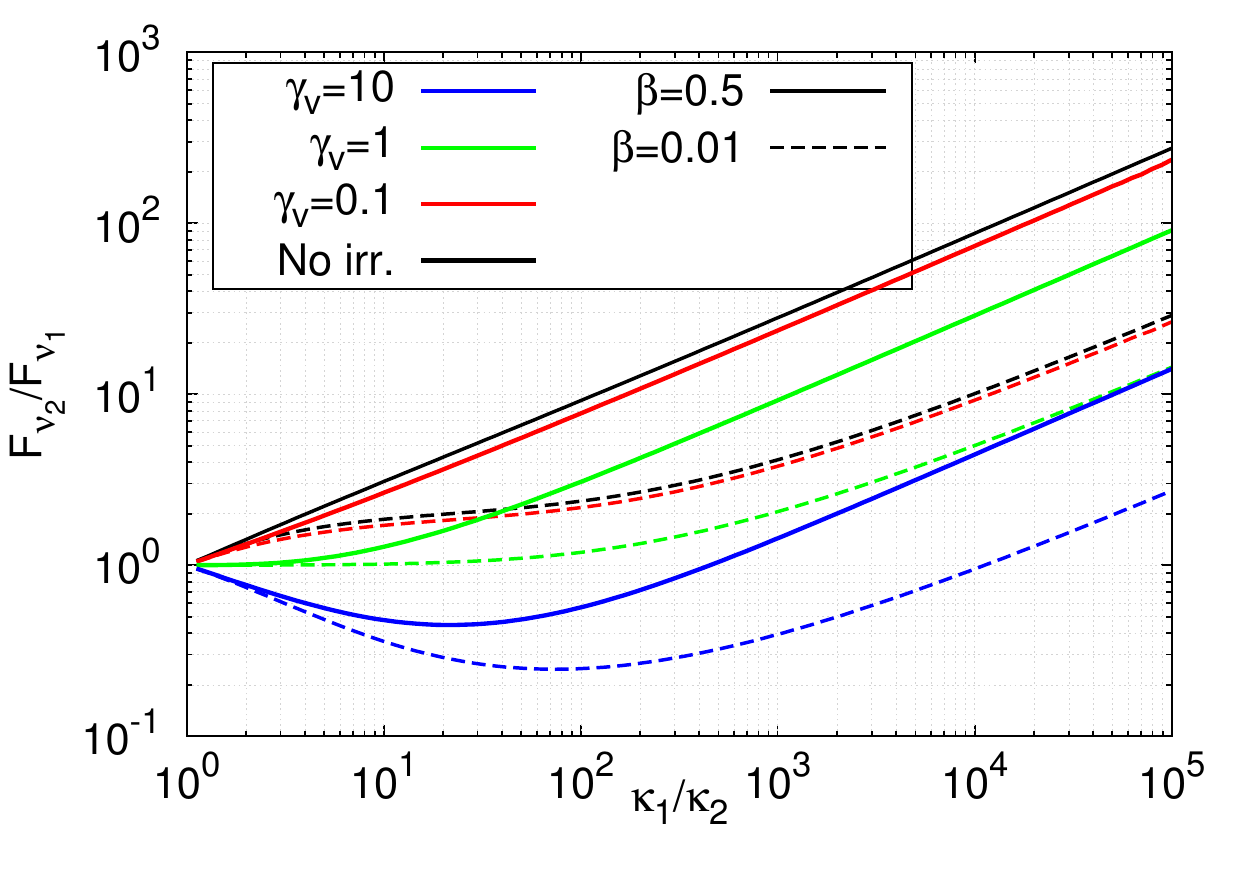}
\caption{Ratio of the monochromatic flux in the two bands $F_{\nu_{\rm 2}}/F_{\nu_{\rm 1}}=\beta H_{2}(0)/(1-\beta)H_{1}(0)$ in function of the opacity ratio $\Ka/\Kb$ for different bandwidth, $\beta$ and visible to infrared opacities, $\Gv$. We used $\mu_{*}=1/\sqrt{3}$}
\label{fig::Fluxes}
\end{figure}

During secondary eclipse observations, the flux emitted by the planet can be observed in different bands \citep[\eg][]{Seager2010}. The detection of molecular species in the emission spectrum of an exoplanet depends strongly on the flux contrast between the continuum and the molecular band considered which depends itself on the temperature profile. Figure \ref{fig::Fluxes} shows the flux per wavelength emitted in the first band ($F_{\nu_{\rm 1}}=4\pi H_{\rm 1}(0)/\beta$) over the flux per wavelength emitted in the second band ($F_{\nu_{\rm 2}}=4\pi H_{\rm 2}(0)/(1-\beta)$). This would be the expected contrast in the emission spectrum of the planet between the spectral features and the continuum. For a non-irradiated atmosphere and for small values of $\Gv$ this is a monotonic function of the opacity ratio $\Ka/\Kb$. The flux in the band of lowest opacity is always bigger than the flux in the band of highest opacity \ie~we see absorption bands. For large values of $\Gv$, whenever a strong inversion happens, the absorption bands turn into emission bands. In any case, for large values of $\Ka/\Kb$ we have:
\begin{equation}
\frac{F_{\nu_{\rm 2}}}{F_{\nu_{\rm 1}}}\propto\left(\frac{\Ka}{\Kb}\right)^{1/2}
\label{eq::FluxOut}
\end{equation}

\section{Resulting temperature profiles}

No matter how strong the non-greyness of the opacities is, there is always a region, at high enough optical depth, where the non-grey solution converges toward the grey solution (see Fig. \ref{fig::Chand} in the non-irradiated case). The transition between a regime where the grey model is accurate to a regime where the non-grey effects are of primordial importance is set by the parameter $\taulim$. For optical depths lower than $\taulim$, non-grey effects are always important, whereas for optical depth higher than $\taulim$, non-grey effects are present only if $\Gv\taulim<1$ and $\beta\to1$. Three distinct situations can be observed on Fig. \ref{fig::TauLim}. For narrow lines ($\beta<0.1$), $\taulim$ is always smaller than one, for larger lines ($0.1<\beta<0.9$), $\taulim$ is close to one whereas for inverted lines ($0.9<\beta<1$), $\taulim$ can reach much bigger values. Thus, when $\Gv>>1$, few non-grey effects are expected in the deep atmosphere, contrary to the cases $\Gv\approx1$ and $\Gv<<1$.

\begin{figure}[!htb]
\includegraphics[width=\linewidth]{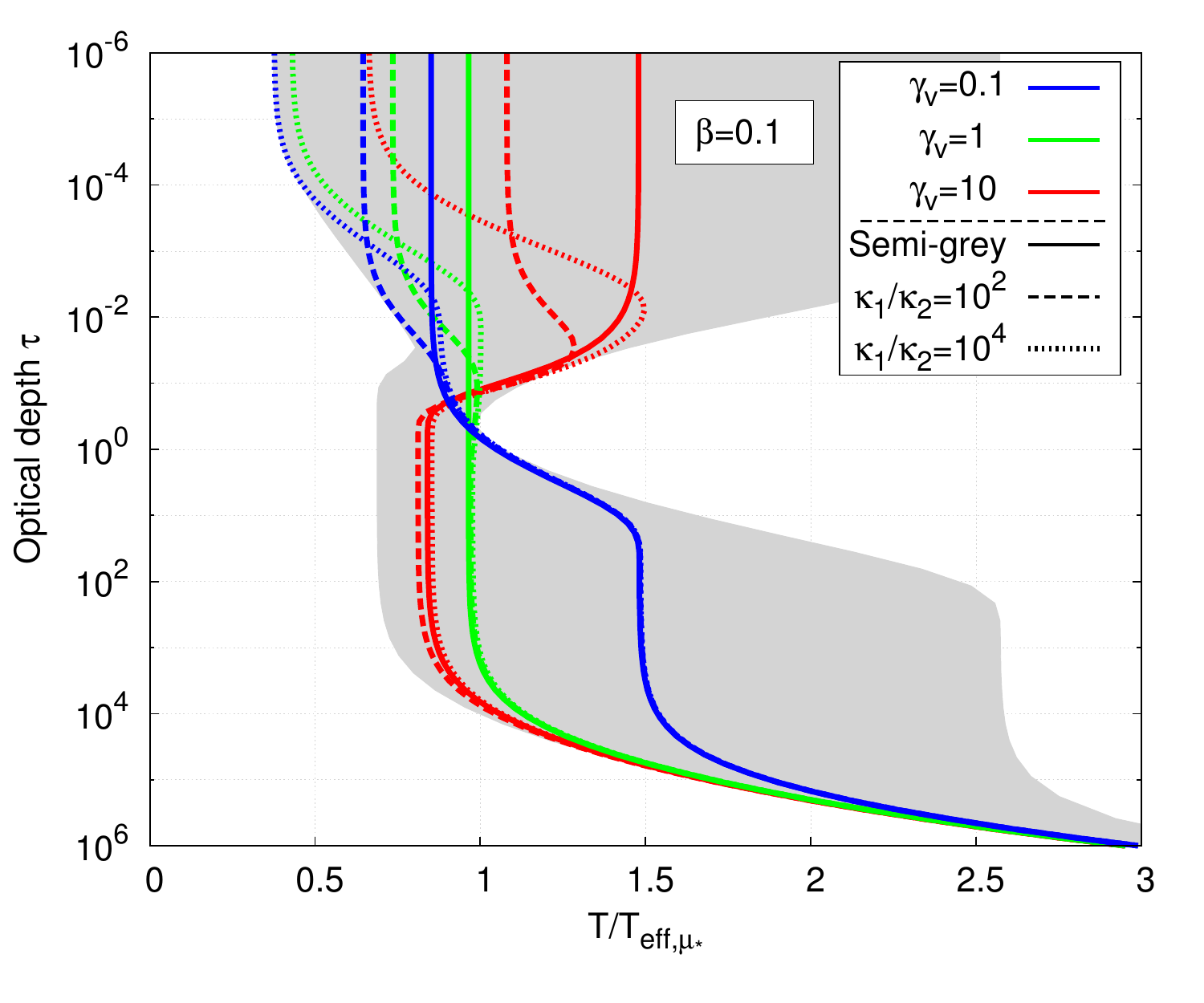}
\caption{PT profiles for an irradiated planet ($\tint=\tirr/10$ and $\mu_{*}=1/\sqrt{3}$). The shaded area show the full range of parameters $10^{-3}<\beta<10^{-1}$, $1<R<10^{4}$ and $0.01<\Gv<100$. The lines are profiles obtained for $\beta=0.1$ and $R=1$ (plain lines), $R=100$ (dashed lines), $R=10^{4}$ (dotted lines) and for $\Gv=0.1$ (blue), $\Gv=1$ (green) and $\Gv=10$ (red).}
\label{fig::AllProfiles1}
\end{figure}

In the case of narrow lines ($\beta<0.1$), only the non-grey cooling of the upper atmosphere is effective. As shown in Fig.~\ref{fig::AllProfiles1}, the profile remains close to the semi-grey model at large optical depth. However at low optical depth, for $\tau<\taulim$, the atmosphere can be much cooler than in the semi-grey case (case $R=1$). In particular, in the $\Gv=10$ case, the non-grey cooling localize the temperature inversion to a specific layer, contrary to the semi-grey case where it extends to the top of the atmosphere. The envelope of all the profiles (shaded area) is much wider than in the semi-grey case (see Fig.\ref{fig::Profiles-G-NG}).

\begin{figure}[!htb]
\includegraphics[width=\linewidth]{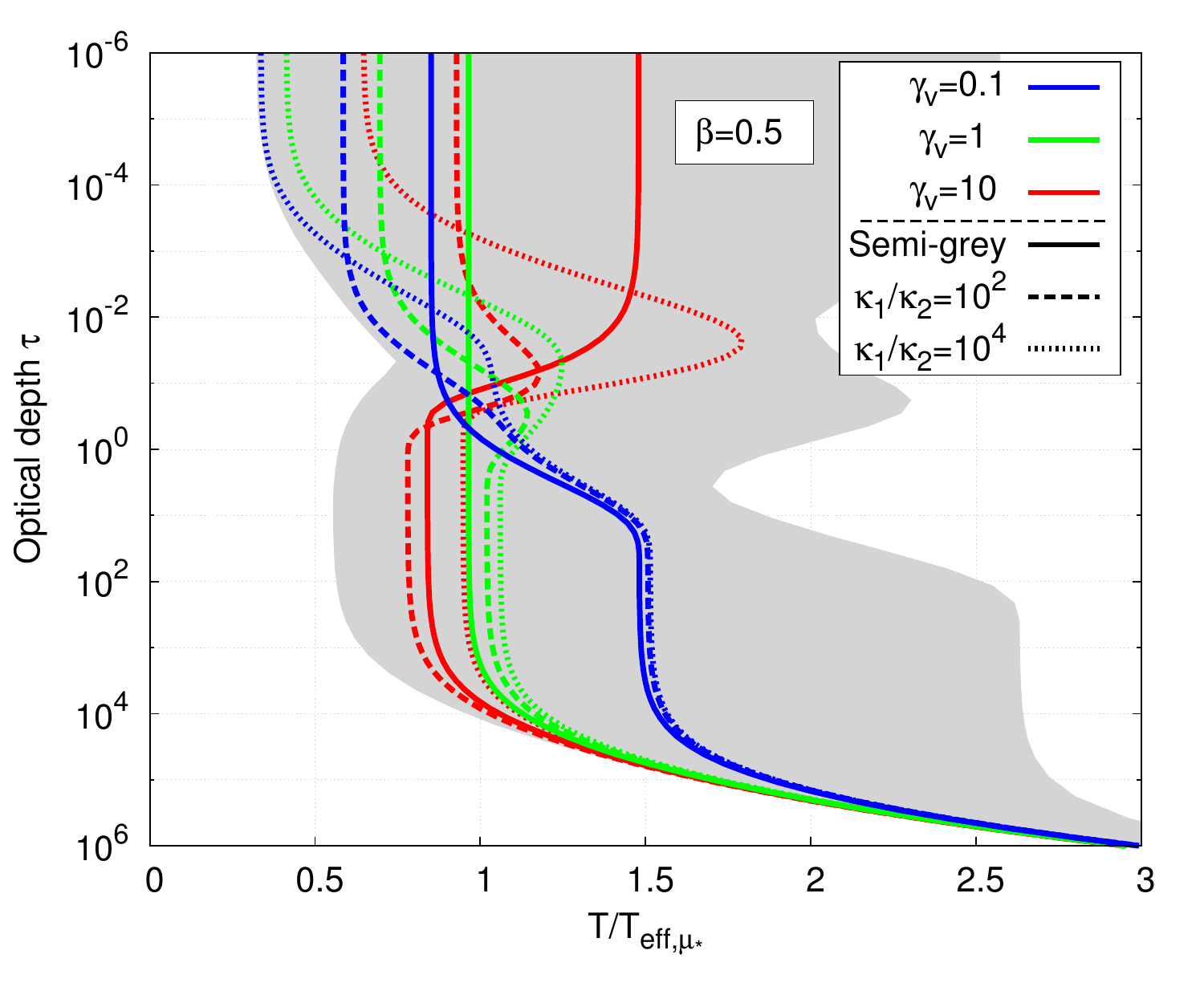}
\caption{PT profiles for an irradiated planet ($\tint=\tirr/10$ and $\mu_{*}=1/\sqrt{3}$). The shaded area show the full range of parameters $0.1<\beta<0.9$, $1<R<10^{4}$ and $0.01<\Gv<100$. The lines are profiles obtained for $\beta=0.5$ and $R=1$ (plain lines), $R=100$ (dashed lines), $R=10^{4}$ (dotted lines) and for $\Gv=0.1$ (blue), $\Gv=1$ (green) and $\Gv=10$ (red).}
\label{fig::AllProfiles2}
\end{figure}

\begin{figure}[!htb]
\includegraphics[width=\linewidth]{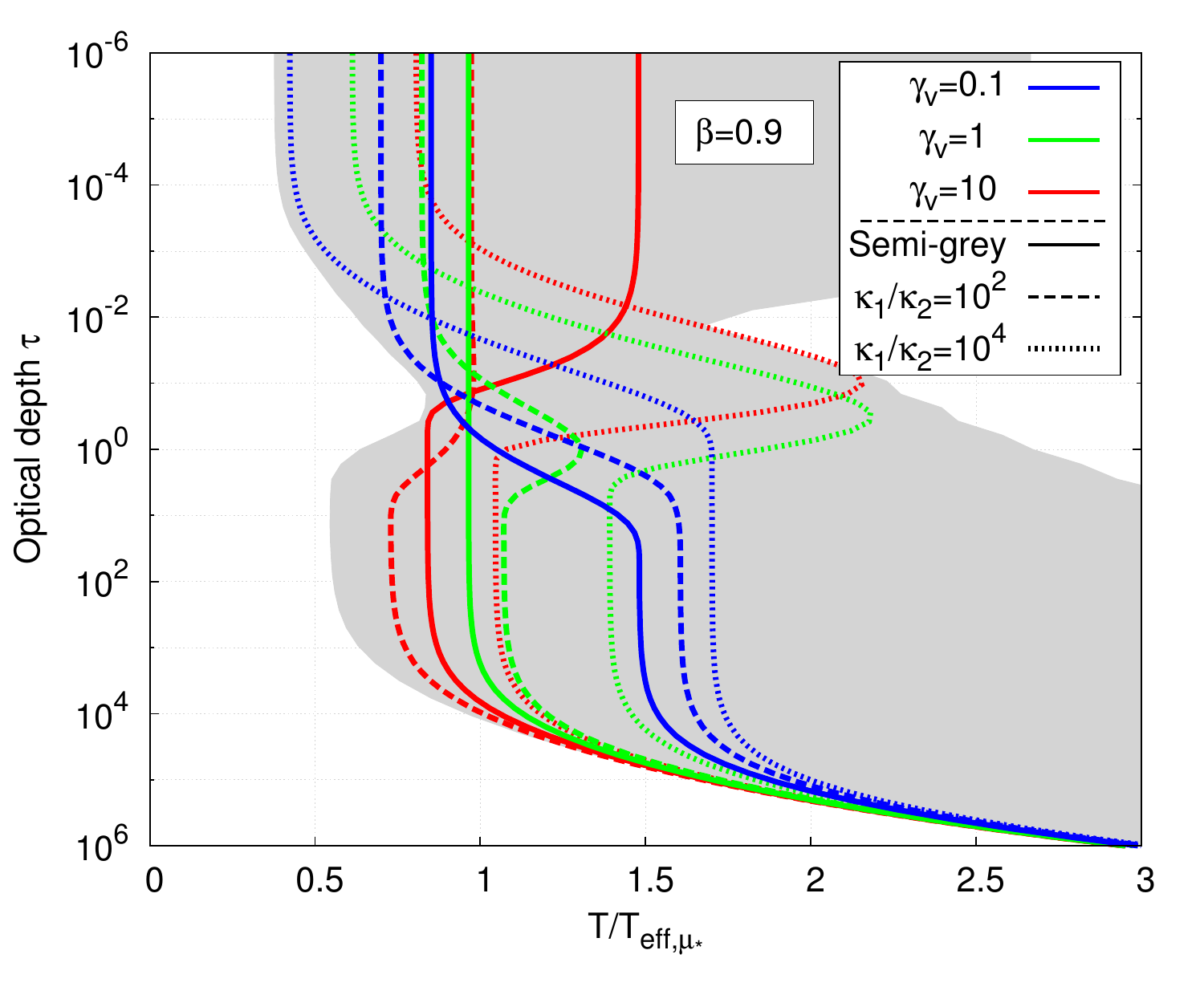}
\caption{PT profiles for an irradiated planet ($\tint=\tirr/10$ and $\mu_{*}=1/\sqrt{3}$). The shaded area show the full range of parameters $10^{-3}<1-\beta<10^{-1}$, $1<R<10^{4}$ and $0.01<\Gv<100$. The lines are profiles obtained for $\beta=0.9$ and $R=1$ (plain lines), $R=100$ (dashed lines), $R=10^{4}$ (dotted lines) and for $\Gv=0.1$ (blue), $\Gv=1$ (green) and $\Gv=10$ (red).}
\label{fig::AllProfiles3}
\end{figure}

In the case of large lines or molecular bands ($0.1<\beta<0.9$) shown in Fig.~\ref{fig::AllProfiles2}, both the non-grey cooling of the upper atmosphere and the blanketing effect are important. Whereas the upper atmosphere undergoes an efficient cooling, the lower atmosphere ($\tau>1$) can experience a significant warming via the blanketing effect. Lowering the ability of the deep atmosphere to cool down efficiently can affect significantly the evolution of the planet \citep{Parmentier2011a,Budaj2012a,Spiegel2013,Rauscher2013a} and could contribute to the radius anomaly of hot-Jupiters \citep[e.g.][]{Guillot2002, Laughlin2011}. Whenever $\Gv\taulim\approx 1$ the stellar irradiation is deposited at a level where non-grey effects lower the ability of the atmosphere to cool down efficiently. This leads to an efficient and localized warming causing a temperature inversion in the profile at $\tau\approx1/\Gv^{*}$, even when none is expected from the semi-grey model (\ie~even when $\Gv^{*}<1$). This happens, for example, when $\beta\approx0.5$ for $\Gv=10$,  when $\beta\approx0.9$ for $\Gv=1$ and for $\beta\approx0.99$ for $\Gv=0.1$ (see Fig.~\ref{fig::AllProfiles3}).

In the case of inverted lines, ($\beta>0.9$), shown in Fig.~\ref{fig::AllProfiles3}, both the upper temperature and the deep temperature are affected by the non-grey effects. The upper atmosphere cools significantly compared to the semi-grey case. The deep atmosphere can either warm up due to the blanketing effect but, for large values of $\Gv$ it becomes cooler than in the semi-grey case (see the case $\Gv=10$ and $R=100$ of Fig.~\ref{fig::AllProfiles3}). Temperatures as cool as $0.5\teffmu$ can be reached. This is fundamentally different from the semi-grey case where the deep temperature is always larger than $2^{1/4}\teffmu$ (see Fig.~\ref{fig::Grey}).

As $\beta$ increases, $\taulim$ increases and the blanketing effect disappear. Eventually, when $\beta\to1$, the opacities, and thus the profile, become semi-grey again. 

In any case, our irradiated picket-fence model can reach the whole temperature range span by the numerical models (see the shaded area in Figs.~\ref{fig::AllProfiles1} to ~\ref{fig::AllProfiles3}). Our model should therefore be preferred to classical semi-grey models as an approximate solution for the temperature profile of irradiated atmospheres.


\section{Conclusion}
We derived an analytic non-grey model to approximate the structure of a plane-parallel irradiated planetary atmosphere. Our model includes non-grey effects in the form of a comb line-opacity function. It is parametrized by $\Gv$, the ratio of the visible to the infrared Rosseland mean opacities, $\Gp$ the ratio of the Planck to the Rosseland mean thermal opacities and $\beta$, the width of the lines. The model is valid for any functional form of the Rosseland mean opacities, the ones obtained from an opacity table for example. However, it cannot account for both realistic Rosseland mean and Planck mean opacities. Their ratio, $\Gp$ and the width of the lines, $\beta$, must be held constant through the atmosphere. Although the model is limited to two thermal opacity bands, it can take into account any number of visible opacity bands, each band adding two new parameters, the strength of the band $\gamma_{\rm v\it{i}}$ and its width $\beta_{\rm v\it{i}}$.

Our model solves the inability of previous analytical models to reach temperatures as cold as predicted by the numerical calculations. For opacities dominated by strong and narrow lines ($\beta<0.1$), non-grey opacities lead to a colder upper atmosphere but converge toward the grey model at optical depth greater than $\taulim$. For opacities dominated by wide lines, or ``bands'' ($\beta\approx0.5$), non-grey opacities still allow the upper atmosphere to cool down more efficiently but also inhibit the cooling of the lower atmosphere. In that case, a significant warming of the lower atmosphere can happen, down to optical depth much greater than $\taulim$. This planetary \emph{blanketing effect} could contribute to the radius anomaly of hot-Jupiters. 

Temperature inversions that were not predicted by previous analytical models occur whenever $\Gv\taulim\approx1$ due to the interaction between the incoming stellar irradiation and the non-grey thermal opacities. These could have interesting observational consequences.

Finally, our model allows for a much greater range of temperature profiles than other solutions of the radiative transfer equations for irradiated atmospheres. We encourage the community to use it when fast calculations of atmospheric temperature profiles are needed. Given the apparent complexity of the solution, a code is available on the internet (see ww.oca.eu/parmentier/nongrey).
 
\section*{Acknowledgements}
This work was performed in part thanks to a joint Fullbright Fellowship to V.P. and T.G. The whole project would not have been possible without the help and support of Douglas Lin. We also acknowledge Jonathan Fortney and Mark Marley for many useful discussions, and the University of California Santa Cruz for hosting us while this work was carried out.

\newpage

\begin{table}
\centering \ra{1.5} 
\begin{tabular}{>{\centering}m{1cm} >{\centering} m{3.5cm} >{\centering} m{1.5cm} >{\centering} m{1cm} }
\toprule 
Symbol & Quantity  & Definition & Units \\ 
\midrule
$\Kr$ &Rosseland mean opacities & eq.~(\ref{eq::Rosseland})  & $\meter\squared\per\kilo\gram$ \\
$\Kp$ &Planck mean opacities & eq.~(\ref{eq::Planck})  & $\meter\squared\per\kilo\gram$ \\
$\Ka$ & Opacity in the first band & eq.~(\ref{eq::CombOpacities})  & $\meter\squared\per\kilo\gram$ \\
$\Kb$ & Opacity in the second band & eq.~(\ref{eq::CombOpacities})  & $\meter\squared\per\kilo\gram$ \\
$\beta$ & Relative width of the first band & eq.~(\ref{eq::beta})  & $-$ \\
$\Kv$ & Opacity in the visible & eq.~(\ref{eq:Gv}) & $\meter\squared\per\kilo\gram$ \\
$\tau$ & Rosseland optical depth & eq.~(\ref{eq::dtau}) & $-$ \\
$\Rt$ & Opacity ratio $\Ka/\Kb$ & sec.\ref{sec::TemperatureProfile} & $-$ \\

$\Ga$, $\Gb$, $\Gp$, $\Gv$ &  $\Ka/\Kr$, $\Kb/\Kr$, $\Kp/\Kr$, $\Kv/\Kr$ & eqs~(\ref{eq:Gp}) \& (\ref{eq:Gv}) &$-$ \\
$\taulim$ &Limit optical depth & eq.~(\ref{eq::taulim}) & $\hertz$ \\

$\nu$ &Frequency & $-$ & $\hertz$ \\
$\mu$ &cosine of the direction angle $\theta$ & $-$ &$-$ \\
$I_{\mu\nu}$ &Specific intensity at frequency $\nu$ and in the direction $\mu$ &\citet{Chandrasekhar1960}  & $\watt\per\meter\squared\per\hertz\per\steradian$ \\
$J_{\nu}$, $H_{\nu}$, $K_{\nu}$ & First, second and third momentum of the specific intensity & eq.~(\ref{eq:MomentDef}) & $\watt\per\meter\squared\per\hertz$ \\
$4\pi H_{\infty}=\sigma T_{\rm int}^{4}$ &Internal flux from the planet & $-$ & $\watt\per\meter\squared$ \\
$\sigma T_{\rm sub}^4$ &Stellar flux arriving at the substellar point of the planet & $-$ & $\watt\per\meter\squared$ \\
$4\pi H_{\rm v}(0)=\sigma T_{\rm irr}^4$ &Stellar flux that penetrates the modeled atmospheric column & $-$ & $\watt\per\meter\squared$ \\
$T_{\rm eff}$ &Effective temperature of the planet & $-$ & $K$ \\
$\teffmu$ &Effective temperature of the modeled atmosphere & $-$ & $K$ \\
$T_{\rm skin}$ & Temperature at optical depth of zero.  & $-$ & $K$ \\
$T_{\rm deep}$ & Temperature at large optical depth.  & eq.~\eqref{eq::Tdeep} & $K$ \\
\bottomrule 
\end{tabular} 
\caption{Main quantities used in this paper.}
 \end{table}

\bibliography{./Parmentier2013b.bib}
\end{document}